\documentclass[amsmath,amssymb,twocolumn,prb,longbibliography,superscriptaddress]{revtex4-1}
\usepackage{graphicx,bm,times}
\usepackage[utf8]{inputenc}
\usepackage{braket}
\usepackage{amsmath}

\usepackage[breaklinks]{hyperref}
\hypersetup{
    bookmarks=false,
    pdfstartview={FitH},
    colorlinks=true,
    linkcolor=blue,
    citecolor=blue,
    urlcolor=blue
}

\newcommand{\nocontentsline}[3]{}
\newcommand{\tocless}[2]{\bgroup\let\addcontentsline=\nocontentsline#1{#2}\egroup}

\begin{document}

\title{Wavefunctions, electronic localization and bonding properties for correlated materials beyond the Kohn-Sham formalism}

\author{A.~D.~N.~James}
\affiliation{H.\ H.\ Wills Physics Laboratory,
University of Bristol, Tyndall Avenue, Bristol, BS8 1TL, United Kingdom}

\author{E.~I.~Harris-Lee}
\affiliation{H.\ H.\ Wills Physics Laboratory,
University of Bristol, Tyndall Avenue, Bristol, BS8 1TL, United Kingdom}

\author{A.~Hampel}
\affiliation{Center for Computational Quantum Physics, Flatiron Institute, 162 5th Avenue, New York, NY 10010, USA}

\author{M.~Aichhorn}
\affiliation{Institute of Theoretical and Computational Physics, TU Graz, NAWI
Graz, Petersgasse 16, 8010 Graz, Austria}

\author{S.~B.~Dugdale}
\affiliation{H.\ H.\ Wills Physics Laboratory,
University of Bristol, Tyndall Avenue, Bristol, BS8 1TL, United Kingdom}

\date{\today}

\begin{abstract}
Many-body theories such as dynamical mean field theory (DMFT) have enabled the description of the electron-electron correlation effects that are missing in current density functional theory (DFT) calculations. However, there has been relatively little focus on the wavefunctions from these theories. We present the methodology of the newly developed {\sc Elk}-TRIQS interface and how to calculate the DFT with DMFT (DFT+DMFT) wavefunctions, which can be used to calculate DFT+DMFT wavefunction dependent quantities. We illustrate this by calculating the electron localization function (ELF) in monolayer SrVO$_3$ and CaFe$_2$As$_2$, which provides a means of visualizing their chemical bonds. Monolayer SrVO$_3$ ELFs are sensitive to the charge redistribution between the DFT, one-shot DFT+DMFT and fully charge self-consistent DFT+DMFT calculations. In both tetragonal and collapsed tetragonal CaFe$_2$As$_2$ phases, the ELF changes weakly with correlation induced charge redistribution of the hybridized As-p and Fe-d states. Nonetheless, the interlayer As-As bond in the collapsed tetragonal structure is robust to the changes at and around the Fermi level.
\end{abstract}

\maketitle

\section{Introduction}

There has been much progress over recent decades in improving upon the electron-electron correlation effects beyond the local density approximation (LDA) \cite{perdew1992} and generalised gradient approximation (GGA) \cite{perdew1996,PBE1996} in density functional theory (DFT).  In the DFT formalism, this includes developments of the meta-GGA functionals \cite{kurth1999,sun2015} and hybrid functionals \cite{Perdew_1996_3,Hyed_2003}. Further progress has also been achieved by combining many-body techniques with DFT, such as DFT with dynamical mean field theory (DFT+DMFT) \cite{Anisimov1997,litchenstein1998,aichhorn2009}. DMFT is successful in describing strongly correlated materials and physical phenomena like the Kondo problem and the Mott transition \cite{georges1996}. These phenomena are absent in current DFT implementations, but are captured in DFT+DMFT. During the development of DFT+DMFT in real materials, there has been an emphasis on comparisons of the calculated quantities with spectroscopic experimental data, such as photoemission spectroscopy (PES), and its angle-resolved version (ARPES), to investigate how the theoretical improvements change the level of agreement between experiment and theory. However, deeper insight into the effect of including many-body techniques will be achieved by calculating the wavefunctions and quantities which depend on them. An example of a wavefunction dependent quantity is the electron localization function (ELF) \cite{becke1990,Savin_1997}, which is often used to help visualize chemical bonds in materials. While it is not possible to directly probe this quantity experimentally, the ELF has nonetheless been used to help understand many materials ranging from molecules \cite{Silvi2005,Santos_2005,Nalewajski_2005} to crystalline structures \cite{kohout2002electron,Cuervo-Reyes_2014,burnus2005}. Therefore, the electron-electron correlation effects that are included from many-body techniques may have significant effects on the ELF which will give further information about the consequences of these interactions.

DFT calculations use an auxillary basis set of fictitious independent single particle Kohn-Sham wavefunctions. When introducing interaction terms, which couple the states of these wavefunctions, it is possible to determine a new basis set in which the wavefunctions would be orthonormal to each other by diagonalizing the charge density matrix. These DFT+DMFT wavefunctions are determined by a basis transformation from the Kohn-Sham basis using a unitary matrix that is derived from diagonalization of the density matrix. Therefore, the DFT+DMFT wavefunctions can be used to calculate wavefunction dependent quantities, which was implemented and is discussed in this paper. Wavefunction dependent quantities also require the electron occupation of states. At zero temperature, these occupations are either zero or one in DFT, whereas in many-body techniques these occupations can have values between zero and one with a discontinuous quasiparticle renormalization at the Fermi level. 
Therefore, changes in the Fermi-liquid-like occupations at the Fermi level or charge redistribution away from the Fermi level will influence wavefunction dependent quantities, especially those in which the correlations are significant. It is therefore imperative to study the effect on the wavefunctions when using many-body techniques in combination with DFT.

This paper introduces the newly developed interface between the full potential linearized augmented plane wave (FP-LAPW) {\sc Elk} \cite{elk} and TRIQS \cite{triqs} open source codes. This {\sc Elk}-TRIQS interface calculates the commonly used spectral DFT+DMFT quantities and gives the opportunity to calculate  wavefunction dependent quantities in {\sc Elk} such as the ELF. See the full interface documentation in Ref. \onlinecite{DFTTools} for further information. We illustrate the usefulness of examining this ELF in two materials - monolayer SrVO$_3$ and bulk CaFe$_2$As$_2$. First we start by discussing the formalism of the interface between the {\sc Elk} package and TRIQS library. This includes the method of calculating the DFT+DMFT wavefunctions from the DFT+DMFT charge density matrix. Then we briefly discuss how the ELF is calculated with these wavefunctions. To benchmark the {\sc Elk}-TRIQS interface, one-shot (OS) and fully charge self-consistent (FCSC) DFT+DMFT calculations were performed on monolayer SrVO$_3$ using both the {\sc Elk}-TRIQS and WIEN2k-TRIQS \cite{aichhorn2016} approaches, with excellent agreement shown between the two. The {\sc Elk}-TRIQS monolayer SrVO$_3$ ELFs are shown to be sensitive to the charge redistribution. Finally, using the {\sc Elk}-TRIQS setup,  we consider the tetragonal (TET) and collapsed tetragonal (CT) phases of CaFe$_2$As$_2$ where an interlayer As-As bond forms in the CT structure, which our ELF calculations also indicate. However, there are slight differences between the ELFs from the GGA, meta-GGA, and FCSC DFT+DMFT approaches that have been used. This is a consequence of the charge redistribution from the hybridized As-p and Fe-d states.

\section{Elk-TRIQS Interface}

The interface from {\sc Elk} to TRIQS consists of {\sc Elk} generating Wannier projectors \cite{Lechermann2006,amadon2008,aichhorn2009dynamical} in a correlated energy window ($\mathcal{W}$), in a similar manner to the procedures described for the WIEN2k interface \cite{aichhorn2016}. The interface from TRIQS back to {\sc Elk} involves the output of the DFT+DMFT density matrix from TRIQS, which is to be read into {\sc Elk} and used to update the electron density, wavefunctions and occupations. The {\sc Elk}-TRIQS interface will be explicitly discussed here. The current implementation of the interface has been tested for non-magnetic, magnetic and non-collinear spin-orbit coupled systems calculated using {\sc Elk}. We are able to interface non-collinear calculations to TRIQS with the potential of doing non-collinear DMFT calculations within TRIQS. For clarity, we will not discuss interfacing spin-orbit coupled and non-collinear systems here. 
 Further information about the interface specifics can be found in Ref. \onlinecite{DFTTools}.

{\sc Elk} uses the second variational method (as described in Ref. \onlinecite{Kurz_2004}) to be able to construct spinors for calculations involving non-collinear magnetism and spin-orbit coupling. By default, {\sc Elk} uses the APW+lo method, but can extend this by including higher order radial differentials in the APWs and local-orbitals. The projectors are constructed directly from the {\sc Elk} Kohn-Sham wavefunctions, meaning that they can be calculated from whatever order APWs and local-orbitals are used in the calculation. In order to construct the Wannier projectors $P_{m \nu}^{\alpha, \sigma}(\mathbf{k})$,  a set of temporary projectors $\widetilde{P}_{m \nu}^{\alpha, \sigma}(\mathbf{k})$ are constructed from the {\sc Elk} Kohn-Sham second variational wavefunctions $\left| \psi_{\mathbf{k} \nu}^{\sigma}\right\rangle$. Here, $\mathbf{k}$, $m$, $\nu$, $\alpha$ and $\sigma$ relate to the $\mathbf{k}$-point, angular momentum $m$ eigenvalue, band index, atom index and spin index, respectively. In order to construct $\widetilde{P}_{m \nu}^{\alpha, \sigma}(\mathbf{k})$, a local basis $| \tilde{\chi}_{m}^{\alpha,\sigma}\rangle$ is used. By default in the {\sc Elk} interface, the local basis is chosen to be the APW radial function ($| \tilde{\chi}_{m}^{\alpha,\sigma}\rangle=\left|u_{l}^{\alpha,\sigma}\left(E_{1 l}\right) Y_{m}^{l}\right\rangle$ with $l$ and $Y_{m}^{l}$ corresponding to the angular momentum and spherical harmonics respectively) within the muffin-tin at the corresponding linearization energy $E_{1 l}$. The temporary projectors are directly computed from the Kohn-Sham second variational wavefunctions by 
\begin{equation}
\widetilde{P}_{m \nu}^{\alpha, \sigma}(\mathbf{k})=\left\langle\tilde{\chi}_{m}^{\alpha,\sigma} | \psi_{\mathbf{k} \nu}^{\sigma}\right\rangle, \quad \nu \in \mathcal{W},
\end{equation}
where $\mathcal{W}$ is the correlated energy window. These temporary projectors are then orthonormalized via
\begin{equation}
P_{m \nu}^{\alpha, \sigma}(\mathbf{k})=\sum_{\alpha^{\prime} m^{\prime}}\left\{[O(\mathbf{k}, \sigma)]^{-1 / 2}\right\}_{m, m^{\prime}}^{\alpha, \alpha^{\prime}} \widetilde{P}_{m^{\prime} \nu}^{\alpha^{\prime}, \sigma}(\mathbf{k}), \label{Orth}
\end{equation}
to form the complete Wannier projector set. Here, $O_{m, m^{\prime}}^{\alpha, \alpha^{\prime}}(\mathbf{k}, \sigma)$ are the overlap matrix elements (equaling to $\langle\tilde{\chi}_{\mathbf{k} m}^{\mathbf{\alpha}, \sigma} | \tilde{\chi}_{\mathbf{k} m^{\prime}}^{\alpha^{\prime}, \sigma}\rangle$) which, in terms of the temporary projectors, have the form
\begin{equation}
O_{m, m^{\prime}}^{\alpha, \alpha^{\prime}}(\mathbf{k}, \sigma)=\sum_{\nu \in \mathcal{W}} \widetilde{P}_{m \nu}^{\alpha, \sigma}(\mathbf{k}) \widetilde{P}_{\nu m^{\prime}}^{\alpha^{\prime}, \sigma *}(\mathbf{k}).
\end{equation}
These Wannier projectors, generated in {\sc Elk}, are read into the TRIQS library along with the energy eigenvalues, symmetries and so on. The treatment of symmetries within this interface is discussed in the Appendix. With these Wannier projectors, we can now calculate the quantities required for the DMFT cycle, which is described in more detail in Ref. \onlinecite{aichhorn2009dynamical}.

For FCSC DFT+DMFT calculations, the wavefunctions, occupations and electron density ($\rho(\mathbf{r})$) within {\sc Elk} are updated from the DMFT outputs by the following method. Starting from the converged OS DFT+DMFT, the self-energy in the Bloch basis, $\Sigma^{\sigma}_{\nu \nu^{\prime}}\left(\mathbf{k}, i \omega_{n}\right)$, is calculated by correcting the impurity self-energy with the double counting correction, i.e.
\begin{equation}
\Delta \Sigma_{m m^{\prime}}^{\sigma, \mathrm{imp}}\left(i \omega_{n}\right)=\Sigma_{m m^{\prime}}^{\sigma, \mathrm{imp}}\left(i \omega_{n}\right)-\Sigma_{m m^{\prime}}^{\mathrm{dc}},
\end{equation} 
which is then upfolded via
\begin{equation}
\begin{array}{ll}
\Sigma^{\sigma}_{\nu \nu^{\prime}}\left(\mathbf{k}, i \omega_{n}\right)=& \\ \sum_{\alpha m m^{\prime}} 
  P_{\nu m}^{\alpha \sigma *}(\mathbf{k})  \Delta \Sigma_{m m^{\prime}}^{\sigma, \operatorname{imp}}\left(i \omega_{n}\right) 
P_{m^{\prime} \nu^{\prime}}^{\alpha, \sigma}(\mathbf{k}).
\end{array}
\end{equation} 
From this, the lattice Green's function $G^{\sigma}_{\nu \nu^{\prime}}\left(\mathbf{k}, i \omega_{n}\right)$ is calculated by 
\begin{equation}
G^{\sigma}_{\nu \nu^{\prime}}\left(\mathbf{k}, i \omega_{n}\right)^{-1}=\left(i \omega_{n}+\mu-\epsilon^{\sigma}_{\mathbf{k} \nu}\right) \delta_{\nu \nu^{\prime}}-\Sigma^{\sigma}_{\nu \nu^{\prime}}\left(\mathbf{k}, i \omega_{n}\right), \label{latG}
\end{equation}
which is used to calculate the interacting charge density matrix $N_{\nu \nu^{\prime}}^{\mathbf{k}, \sigma}$. This is calculated from the summation of $G^{\sigma}_{\nu \nu^{\prime}}\left(\mathbf{k}, i \omega_{n}\right)$ over the Matsubara frequencies,
\begin{equation}
N_{\nu \nu^{\prime}}^{\mathbf{k},\sigma}=\sum_{n} G^{\sigma}_{\nu \nu^{\prime}}\left(\mathbf{k}, i \omega_{n}\right) e^{i \omega_{n} 0^{+}}, \quad \nu \in \mathcal{W}. \label{Nvv}
\end{equation}
Then $N_{\nu \nu^{\prime}}^{\mathbf{k}, \sigma}$ is read into {\sc Elk}.

In general, $\Sigma^{\sigma}_{\nu \nu^{\prime}}\left(\mathbf{k}\right)$ has non-diagonal elements which consequently means that $N_{\nu \nu^{\prime}}^{\mathbf{k},\sigma}$ would also have non-diagonal elements. To construct a diagonal set of wavefunctions (as is conventionally used in DFT), first the total density matrix ($N_{\nu \nu^{\prime}}^{\prime \mathbf{k},\sigma}$) within the DFT Kohn-Sham basis is constructed by combining the DMFT density matrix within the correlated energy window $\mathcal{W}$ ($N_{\nu \nu^{\prime}}^{\mathbf{k},\sigma}$, $\nu$ $\in$  $\mathcal{W}$) with the DFT density matrix outside that window ($n_{\nu \nu^{\prime}}^{\mathbf{k},\sigma}$, $\nu$ $\notin$ $\mathcal{W}$). This total density matrix includes all the DFT+DMFT state indices $\nu$, both within and outside $\mathcal{W}$. A new set of diagonal DFT+DMFT occupation numbers $\mathcal{N}_{\zeta}^{\mathbf{k},\sigma}$, and DFT+DMFT wavefunctions $| \phi_{\mathbf{k} \zeta}^{\sigma}\rangle$ are determined by a unitary transformation $U$ given by diagonalizing the total density matrix (as in Ref.~\onlinecite{schuler2018}):

\begin{equation}
\mathcal{N}_{\zeta}^{\mathbf{k},\sigma} \delta_{\zeta\zeta\prime} = \sum_{\nu \nu^{\prime}} U^{\sigma}_{\zeta\nu}N_{\nu \nu^{\prime}}^{\prime \mathbf{k},\sigma}U^{\sigma *}_{\nu\prime\zeta\prime},
\end{equation}
\begin{equation}
| \phi_{\mathbf{k} \zeta}^{\sigma}\rangle = \sum_{\nu} U^{\sigma}_{\zeta\nu} | \psi_{\mathbf{k} \nu}^{\sigma}\rangle,
\end{equation}

where $\delta_{\zeta\zeta\prime}$ is the kronecker delta.
 These DFT+DMFT wavefunctions can be used to generate quantities that are solely dependent on the wavefunctions and occupations. It is these DFT+DMFT wavefunctions and occupations that we have used to generate the ELF in subsequent sections. The FCSC DFT+DMFT cycle uses these wavefunctions and occupations to calculate the new DFT+DMFT electron density, $\rho(\mathbf{r})$, which is now
\begin{equation}
\rho(\mathbf{r})=\sum_{\zeta,\mathbf{k},\sigma} \mathcal{N}_{\zeta}^{\mathbf{k},\sigma}  \langle \mathbf{r} | \phi_{\mathbf{k} \zeta}^{\sigma}\rangle \langle \phi_{\mathbf{k} \zeta}^{\sigma} | \mathbf{r} \rangle.
\end{equation}

The density matrix includes information about both occupied incoherent and coherent quasiparticle states as these are included in the Matsubara lattice Green’s function, and the density matrix is derived from the summation of this Green’s function in Eq.~\ref{Nvv}. This is equivalent to integrating the occupied spectral function on the real frequency axis but note that it does not require any analytic continuation which complicates that approach. The influence of these coherent and incoherent states changes the occupation function (derived from the density matrix) from a Fermi-Dirac function in DFT to something more like the Fermi-Liquid occupation function.

Finally, the total energy in the FCSC DFT+DMFT formalism is calculated the same way as in previous implementations \cite{aichhorn2016,schuler2018}.  From the electron density update, a new set of {\sc Elk} Kohn-Sham second variational wavefunctions can be generated from $\rho(\mathbf{r})$ (by solving the Kohn-Sham equations once) to produce new Wannier projectors and hence complete an FCSC DFT+DMFT cycle.

\section{Electron Localization Function}
In the previous section, we have shown how to calculate the DFT+DMFT wavefunctions and occupation numbers without the need of analytic continuation. These can be used to calculate the ELF which is dependent on the occupations and wavefunctions of the system. The ELF is based on a same-spin pair probability density $D(\mathbf{r})$ of finding an electron close to another same-spin reference electron \cite{becke1990}. It has mainly been used as a tool to investigate the electron localization in chemical bonds via Hartree-Fock \cite{becke1990} and DFT \cite{Savin_1992} methods. The ELF has also been used to investigate the bond evolution in time-dependent DFT (TDDFT) \cite{burnus2005}. The ELF has the form 
\begin{equation}
\eta(\mathbf{r})=\frac{1}{1+\left[D(\mathbf{r}) / D^{0}(\mathbf{r})\right]^{2}},  \label{e:elf}
\end{equation}
where
\begin{equation}
D^{0}(\mathbf{r})=\frac{3}{5}\left(6 \pi^{2}\right)^{2 / 3}\left(\frac{\rho(\mathbf{r})}{2}\right)^{5 / 3}
\end{equation}
is the kinetic energy density for the homogeneous electron gas as a function of the electron density ($\rho(\mathbf{r})$), and $D(\mathbf{r})$ is given by
\begin{equation}
D(\mathbf{r})=\frac{1}{2}\left(\tau(\mathbf{r})-\frac{1}{4} \frac{[\nabla \rho(\mathbf{r})]^{2}}{\rho(\mathbf{r})}\right),
\end{equation}
with $\tau(\mathbf{r})$ being the spin-averaged kinetic energy density from the wavefunctions, which has the form
\begin{equation}
\tau(\mathbf{r})=\sum_{\mathbf{k} i}n_{\mathbf{k} i}\left|\nabla \Psi_{\mathbf{k} i}(\mathbf{r})\right|^{2}. 
\end{equation}
The wavefunctions $\Psi_{\mathbf{k} i}(\mathbf{r})$ and occupation numbers $n_{\mathbf{k} i}(\mathbf{r})$ that are used to calculate an ELF can be any diagonal wavefunction (and occupation) set, such as those from the DFT or DFT+DMFT calculations. Hence, index $i$ would refer to index $\nu$ or $\eta$ for the DFT or DFT+DMFT calculation respectively. Therefore, the effect of the electron-electron correlation approximations on the wavefunctions - and consequently on $\rho(\mathbf{r})$ and $\tau(\mathbf{r})$ - can be investigated by using the ELF.

It is evident from Eq.~\ref{e:elf} that the ELF is a quantity which can vary from 0 to 1, with a reference value of 0.5 relating to the Pauli repulsion being equal to that from an homogeneous electron gas with the same density $\rho(\mathbf{r})$. ELF values that tend to 1 relate to a $D(\mathbf{r})$ that tends to zero with respect to the homogeneous electron gas, and therefore the electrons would be highly localized in that region of space. 
It should be noted, however, that a direct relationship between the ELF and the Pauli exclusion of the electrons (i.e. their localized or itinerant nature), is difficult to deduce as the ELF is dependent on $D(\mathbf{r})/D^{0}(\mathbf{r})$ not just the same-spin pair probability density of the material \cite{kohout1997influence,kohout2002electron}. 

The many-body effects from DMFT (encoded in the density matrix and DFT+DMFT wavefunctions) will change the $\rho(\mathbf{r})$ and $\tau(\mathbf{r})$ distributions, which in turn modify both the $D(\mathbf{r})$ and $D^{0}(\mathbf{r})$ on which the ELF depends. The extent of the changes in $\rho(\mathbf{r})$ and $\tau(\mathbf{r})$ will be material specific, so linking the ELF distribution to just one of these may not always be possible.  Therefore, comparing the ELF from different theoretical approaches will give insight into the interplay of the changes to the $\rho(\mathbf{r})$ and $\tau(\mathbf{r})$ distributions as well as the changes to the bonding present in the material. The charge redistribution in the materials studied here dominated the changes in the ELF.

\section{Results}

\subsection{Monolayer SrVO$_3$}

\begin{figure}[h]
\centerline{\includegraphics[width=\linewidth]{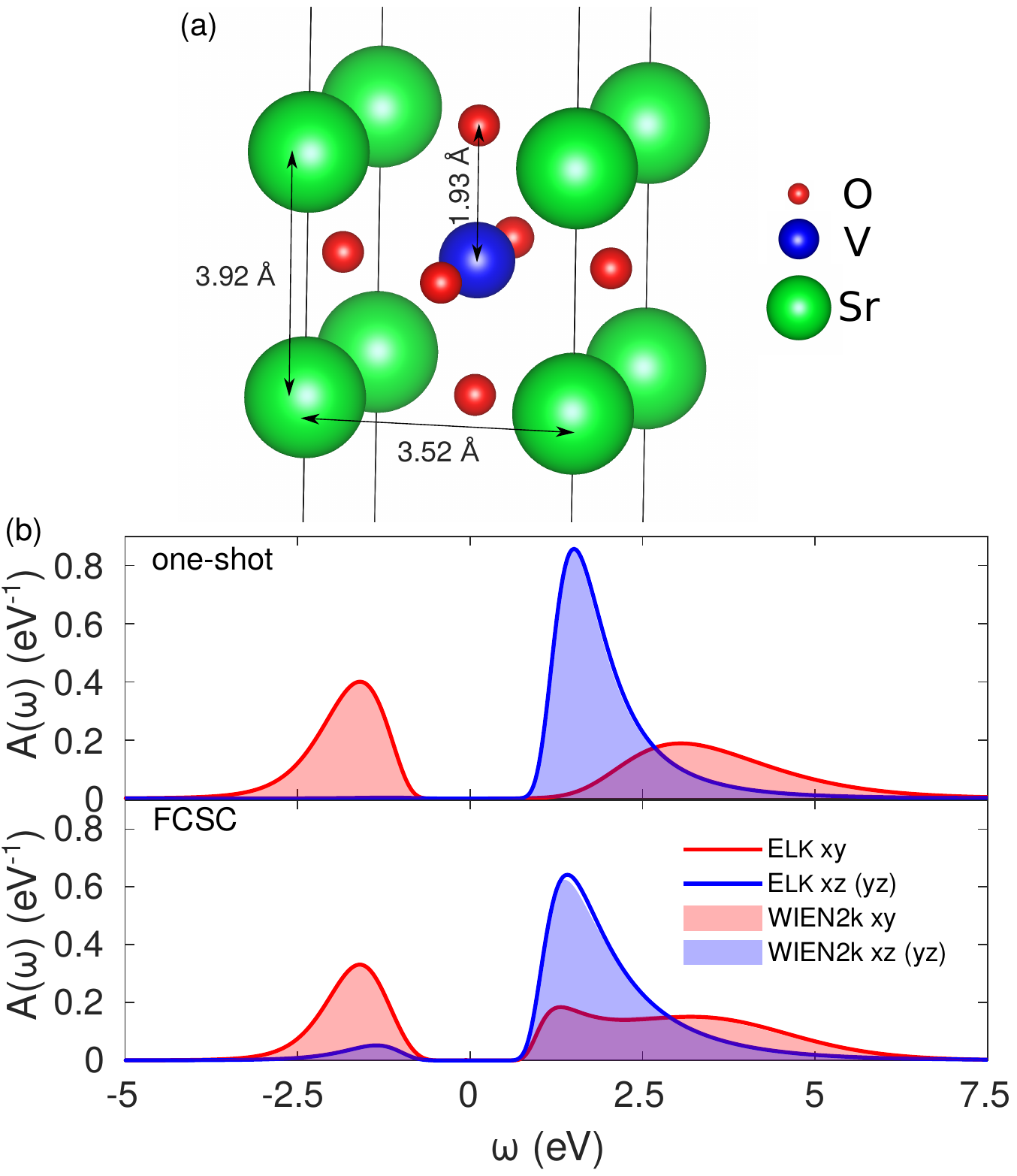}}
\caption{(a) The unit cell of the monolayer SrVO$_3$, where each monolayer is separated by 20${\rm \AA}$ of vacuum. (b) The comparison of the Wannier V t$_{\rm 2g}$ spectral functions calculated in the one-shot and fully charge self-consistent (FCSC) DFT+DMFT methods by the WIEN2k-TRIQS and {\sc Elk}-TRIQS code combinations.}
\label{f:SVO}
\end{figure}

\begin{figure*}[t!]
\centerline{\includegraphics[width=0.85\linewidth]{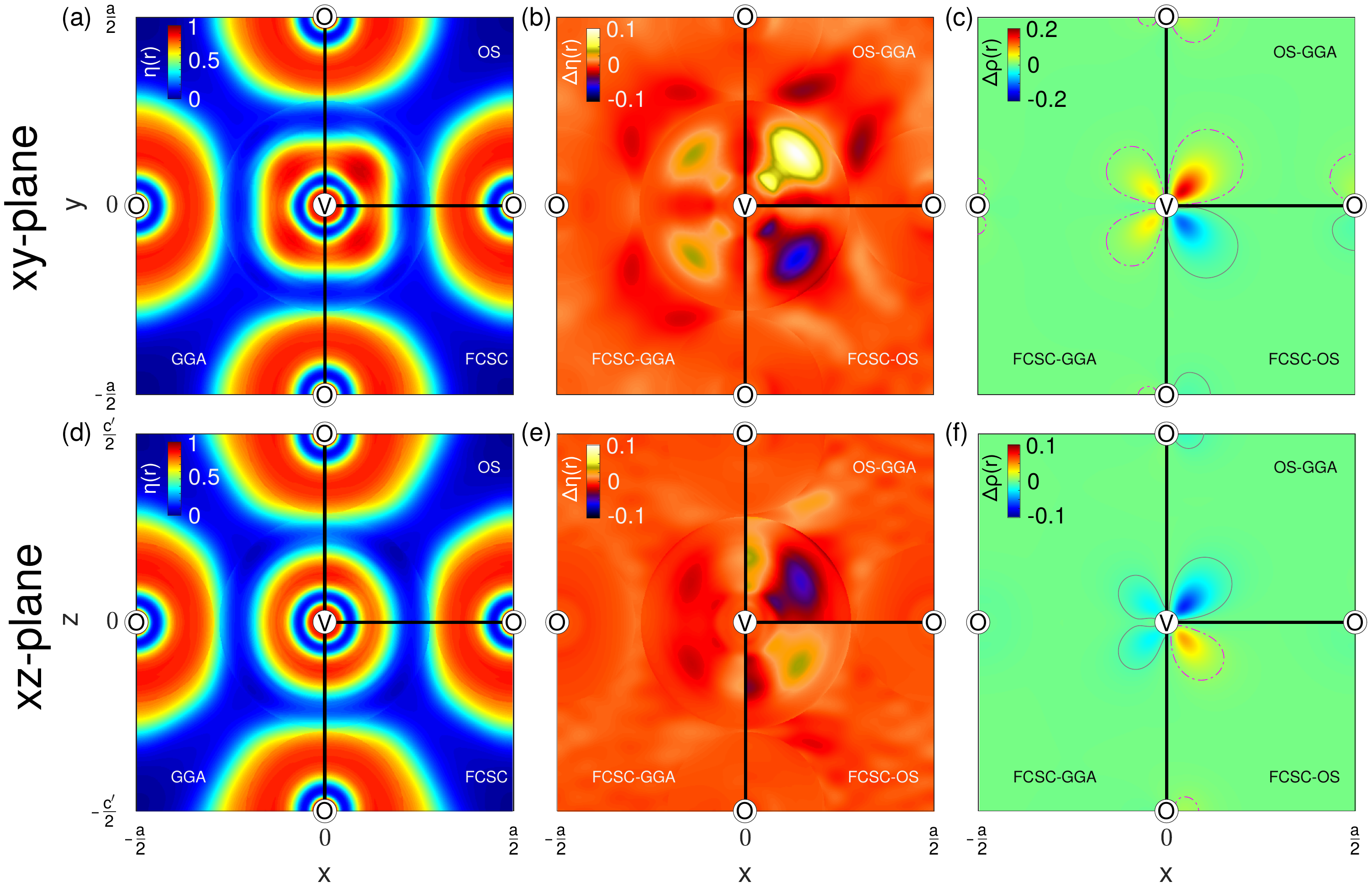}}
\caption{The monolayer SrVO$_3$ (a) xy- and (b) xz-plane ELFs, slicing through the center of the V and O atoms. The four-fold symmetry of the planes has been exploited to show the results from the GGA (PBE), one-shot (OS) and fully charge self-consistent (FCSC) DFT+DMFT calculations. (b) and (e) are the differences (for example FCSC-GGA = $\eta^{\rm FCSC}(\textbf{r})$ - $\eta^{\rm GGA}(\textbf{r})$) in the ELFs from the different theoretical techniques in the xy- and xz-planes respectively. (c) and (f) show the charge density differences (for example FCSC-GGA = $\rho^{\rm FCSC}(\textbf{r})$ - $\rho^{\rm GGA}(\textbf{r})$) between the different theoretical techniques in the xy- and xz-planes respectively. The grey solid and magenta dot-dashed contours in (c) and (f) show the positive and negative charge density differences isovalues of 6$\times$10$^{-3}$. The charge density differences are in units of electrons/Bohr$^3$.}
\label{f:SVO_ELFS}
\end{figure*}

Monolayer SrVO$_3$ is a Mott insulator material in which the charge redistribution is significant between the OS and FCSC DFT+DMFT methods. This material has been used before to benchmark the VASP-TRIQS interface \cite{schuler2018}. Bulk SrVO$_3$ is a prototypical correlated material for which many DFT+DMFT studies \cite{nekrasov2005,georges1996,sekiyama2004,nekrasov2006,kotliar2006,byczuk2007,tomczak2012} have produced good agreement with the experimentally observed three-peak structure, which DFT calculations cannot currently capture. It is the V t$_{\rm 2g}$ states which have been treated as correlated in previous DFT+DMFT calculations, giving the calculated Hubbard states. In bulk SrVO$_3$, often only the V t$_{\rm 2g}$ states around the Fermi level are used in the DMFT calculations. The e$_{\rm g}$ states were not considered as correlated as they do not hybridize with the t$_{\rm 2g}$ states and are also unoccupied. On the other hand, recent GW+DMFT work in Ref. \onlinecite{Boehnke2016} reinterprets SrVO$_3$ as a weakly correlated material with low static local interactions, since their results show pronounced plasmonic satellites due to screening. Reducing the dimensionality of bulk SrVO$_3$ to a monolayer causes a metal-insulator transition (MIT). This is seen experimentally \cite{yoshimatsu2010,gu2014} and complimented by DFT+DMFT calculations \cite{bhandary2016,schuler2018}. There are a few material-specific mechanisms which the MIT has been attributed to such as the crystal field splitting \cite{zhong2015,hampel2020,beck2018,sclauzero2016} and confinement in the SrVO$_3$ layers \cite{james2020quantum}.  

Here, we apply both the OS and FCSC DFT+DMFT methods to the relaxed monolayer SrVO$_3$ calculations using the {\sc Elk}-TRIQS and WIEN2k-TRIQS interfaces. The relaxed structure has been determined by previous GGA calculations \cite{zhong2015}. The monolayer structure is shown in Fig.~$\ref{f:SVO}$ (a). The $a$ and $c$ lattice parameters are 3.52 $\rm \AA$ and 3.92 $\rm \AA$ respectively, where the $a$ lattice parameter is of bulk SrTiO$_3$. The unit cell used in the DFT calculations has a separation of 20 $\rm \AA$ between the monolayers. The out-of-plane V-O distance has increased to 1.93 $\rm \AA$ compared to $c$/2.

The WIEN2k \cite{blaha2020wien2k} and {\sc Elk} DFT calculations used a k-mesh of 15$\times$15$\times$1 and the PBE \cite{PBE1996} GGA functional, which is the same as that used in Ref. \onlinecite{schuler2018}. The correlated V t$_{\rm 2g}$ states around the Fermi level are the subject of the DMFT calculations. As with the bulk, the e$_{\rm g}$ states in the monolayer are unoccupied and do not hybridize with the t$_{\rm 2g}$ states. Therefore, the e$_{\rm g}$ states were not used in DMFT, which is consistent with previous benchmarking calculations \cite{bhandary2016,schuler2018}. The V t$_{\rm 2g}$ Wannier 
projectors were generated within a correlated energy window around the Fermi energy of [-2.0, 1.1] 
eV and then these projectors were interfaced to the TRIQS library \cite{triqs} by the 
TRIQS/DFTTools application \cite{aichhorn2016}. The DMFT calculations used the continuous-time quantum Monte Carlo (CTQMC) solver in the TRIQS/CTHYB application \cite{seth2016} 
with $4.2 \times 10^7$ sweeps and the Hubbard-Kanamori interaction Hamiltonian. The double counting was approximated in the fully localized limit (FLL) which used the DMFT occupations. These DMFT calculations used previously defined  \cite{bhandary2016,schuler2018} $U$~=~5.5~eV, $J$~=~0.75~eV, and inverse temperature $\beta$~=~40~eV$^{-1}$.

\begin{table}[h]
\begin{tabular}{c|cc|cc|cc}
  &\multicolumn{2}{c}{GGA}
  &\multicolumn{2}{c}{OS}
  &\multicolumn{2}{c}{FCSC}   \\
  & d$_{\rm xy}$ & d$_{\rm xz}$+d$_{\rm yz}$ & d$_{\rm xy}$ & d$_{\rm xz}$+d$_{\rm yz}$ & d$_{\rm xy}$ & d$_{\rm xz}$+d$_{\rm yz}$  \\
\hline
\hline
{\sc Elk}~/~WIEN2k & 0.65 & 0.35 & 0.98 & 0.02 & 0.76 & 0.24  \\
\hline
\end{tabular}
\caption{\label{t:SVO_ne} Comparison of the GGA (PBE), one-shot (OS) and fully charge self-consistent (FCSC) DFT+DMFT orbital charges based on the {\sc Elk} and WIEN2k DFT codes.}
\end{table}

Fig.~$\ref{f:SVO}$ (b) shows the spectral function comparison between the {\sc Elk}-TRIQS and WIEN2k-TRIQS DFT+DMFT calculations of both the OS and FCSC methods. These spectral functions were obtained by analytically continuing the DMFT local Green's function by the {\em LineFitAnalyzer} technique of the maximum entropy analytic continuation method implemented within the TRIQS/Maxent application \cite{maxent}.  The ELK-TRIQS and WIEN2k-TRIQS spectral functions are in excellent agreement with each other, with both showing the incoherent Hubbard peaks. However, there are some minor discrepancies between the spectral functions which are mainly a consequence of the ill-posed problem in the analytic continuation process. Both {\sc Elk}-TRIQS and WIEN2k-TRIQS calculations produce the strong orbital charge polarization seen in the OS calculations which is softened in the FCSC results caused by charge redistribution. This redistribution occurs at the DFT level in the FCSC DFT+DMFT cycle when the correlated t$_{\rm 2g}$ states are fed back into DFT \cite{bhandary2016}. These results agree with previous studies of this monolayer \cite{bhandary2016,schuler2018}. The Wannier orbital charges are identical in Table \ref{t:SVO_ne} when quoted to 2 significant figures. There are minor discrepancies between Elk-TRIQS and WIEN2k-TRIQS present in the DFT Wannier charges when quoted to higher significant figures. These discrepancies will have propagated to the DFT+DMFT calculations. These reasonable DFT discrepancies can be explained by the different (default) set of local-orbitals used, the muffin-tin radial functions being evaluated at different linearization energies, as well as the different hard coded approaches the DFT packages used to implement the APW+lo method. A discussion on the comparison of quantities evaluated by different DFT codes, and their discrepancies, can be found in Ref.~\onlinecite{Lejaegherea_2016}.
In the DFT+DMFT calculations on the other hand, the results between the Elk-TRIQS and WIEN2k-TRIQS are the same to 2 significant figures. The values are quoted to this precision because of the inherent noise present from the CTQMC solver which causes the charge to fluctuate in the higher significant figures. It should be noted that the OS orbital charges are slightly different here compared to the results in Ref.~\onlinecite{schuler2018}. This is because more OS DFT+DMFT cycles were done here for better convergence. This discrepancy does not change their conclusions. Nonetheless, the excellent agreement shown in fig.~$\ref{f:SVO}$ (b) gives confidence that the {\sc Elk}-TRIQS interface works.

\begin{figure*}[t]
\centerline{\includegraphics[width=\linewidth]{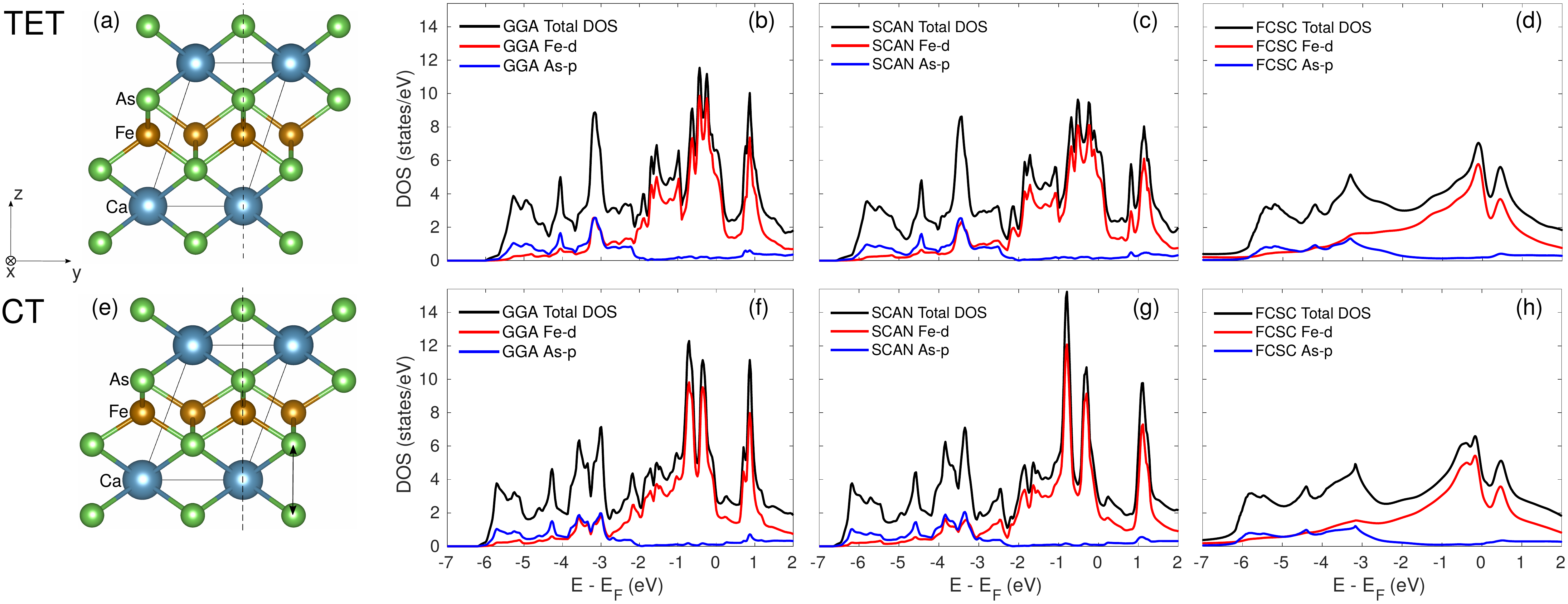}}
\caption{(a) and (e) show the structures of the tetragonal (TET) and collapsed tetragonal (CT) phases, respectively. The parallelepiped Wigner-Seitz unit cell is shown along with the dashed line indicating the xz-plane on which the ELFs were calculated and in (e) the interlayer As-As bond is also indicated by the double arrow. (b), (c) and (d) are total and partial density of states for the GGA (PBE), SCAN and fully charge self-consistent (FCSC) DFT+DMFT calculations in the TET structure. (f), (g) and (h) show the same quantities as (b), (c) and (d), respectively, in the CT structure.}
\label{f:CFA_DOS}
\end{figure*}

As the ELF ($\eta(\textbf{r})$) depends on wavefunction dependent quantities, the effect of the electron-electron correlation effects from the different theoretical methods can be investigated. Fig.~\ref{f:SVO_ELFS} shows $\eta(\textbf{r})$, the difference in $\eta(\textbf{r})$ ($\Delta\eta(\textbf{r})$) and the differences in $\rho(\textbf{r})$ ($\Delta\rho(\textbf{r})$) between the DFT and DFT+DMFT techniques for the V-centered xy- and xz-planes (the yz-plane $\eta(\textbf{r})$ is the same as in the xz-plane due to symmetry). The xy and xz $\eta(\textbf{r})$, in Figs.~$\ref{f:SVO_ELFS}$ (a) and (d), are similar with the greatest values located in shells around the V and O atoms. However, the xy-plane $\eta(\textbf{r})$ has the significantly filled V d$_{\rm xy}$ orbital nodes around the V atom, whereas the d$_{\rm xz(yz)}$ orbital nodes are not present in the xz-plane $\eta(\textbf{r})$ as these orbitals have little charge contribution. The contribution of the V t$_{\rm 2g}$ orbitals to $\eta(\textbf{r})$ in the different DFT and DFT+DMFT calculations can be seen in the $\Delta\eta(\textbf{r})$ of the xy- and xz-planes in Figs.~$\ref{f:SVO_ELFS}$ (b) and (e), respectively. For the d$_{\rm xy}$/d$_{\rm xz}$ orbitals, it can be seen that the OS calculation increases/reduces significant contributions to the $\eta(\textbf{r})$ compared to the DFT. On the other hand, the d$_{\rm xy}$/d$_{\rm xz}$ orbital contributions to the FCSC $\eta(\textbf{r})$ only slightly increases/reduces with respect to the DFT, so that differences between the FCSC and DFT are not so significant. This follows the d$_{\rm xy}$ and d$_{\rm xz}$ $\Delta\rho(\textbf{r})$ orbital contributions from the DFT and DFT+DMFT calculations highlighted in Figs.~$\ref{f:SVO_ELFS}$ (c) and (f), respectively. It should be noted that in Figs.~$\ref{f:SVO_ELFS}$ (b) and (e) the muffin-tin and interstitial regions are distinctly visible which is an artifact of the basis set used.

The ELF in monolayer SrVO3 is sensitive to the redistribution of charge, with respect to DFT, caused by the application of both the OS and FCSC DFT+DMFT methods. In this material, the changes in the ELF can be traced to the changes in the charge distribution (primarily from the t$_{\rm 2g}$ orbitals as expected), indicating that it is that which dominates when DMFT is included. From Figs.~$\ref{f:SVO_ELFS}$ (c) and (f), small changes to the charge redistribution of the O occur even though only the t$_{\rm 2g}$ states are treated as correlated. This is because the Wannier projectors contain information about the hybridized V-t$_{\rm 2g}$ and O-p states (the same as in the bulk - see for example Ref.~\onlinecite{Lechermann2006}). The changes in the charge around the O sites appear to be too small to affect the ELF significantly. The ELF here gives another means to visualize the impact of DMFT in correlation-induced changes on the electron distribution within the bonds in this material.

\subsection{CaFe$_2$As$_2$}

The CaFe$_2$As$_2$ compound is a member of the 122 AFe$_2$As$_2$ (A being an alkaline metal) family of Fe-pnictide superconductors. This material has been reported to have three distinct phases: the TET phase which is the structure that exists at room temperature and ambient pressure; the antiferromagnetic orthorhombic phase; and the CT phase which displays superconductivity under uniaxial pressure \cite{Kreyssig2008,budko2016}. As well as superconductivity, CaFe$_2$As$_2$ has displayed the shape memory alloy and superelasticity effects \cite{sypek2017superelasticity}. Chemical substitution enables the fine tuning of CaFe$_2$As$_2$ properties \cite{Saha2012,knoner2016,Ran2014,Nohara2017,Pobel2013,Ortenzi2015,Yang2015}.
 
ARPES studies have helped to understand the role of correlations in the TET and CT phases \cite{vanroekeghem2016,dhaka2014}. The DFT+DMFT comparisons with the experimental results \cite{vanroekeghem2016,diehl2014correlation,Mandal2014} have improved agreement compared to the DFT calculations. Although the DFT is able to describe some of the ARPES features well, the DFT+DMFT results capture some of the band renormalization. However, the renormalization is still not as significant as seen in the ARPES data which has been suggested to be attributed to both non-local correlations and/or phonon effects \cite{diehl2014correlation}.

Previous studies on AB$_2$X$_2$ compounds (to which CaFe$_2$As$_2$ belongs) show that an interlayer X-X $\sigma$ bond can exist from the X-p states overlapping \cite{hoffmann1985,zheng1988,Nohara2017}. The bonds between the A and B$_2$X$_2$ layers are ionic, the B-B are metal-metal bonds and the B-X and X-X bonds are covalent in nature. The B-X bond is associated to the hybridization of the B-d and X-p states below the Fermi level \cite{hoffmann1985}. X-ray diffraction and Raman spectroscopy supports the existence of the X-X bond in NaFe$_2$As$_2$ \cite{stavrou2015}. In CaFe$_2$As$_2$, the As-As interlayer bond has been shown to be mediated by the Fe-As bond and the Fe spin-state \cite{Yildirim2009}.

As the real-space ELF and charge distribution changes with respect to the inclusion of the electron-electron correlation effects from many-body techniques, the ELF will provide information on the effect of different types of included electron-electron correlation effects on the As-As bond. Therefore, we have calculated the ELF via GGA (PBE \cite{PBE1996}) and meta-GGA functionals, as well as the {\sc Elk}-TRIQS FCSC DFT+DMFT implementation, to determine how the changes in the electron-electron correlation effects between each method affects the ELF distribution around the As atoms in the TET and CT structures, along with the consequences this has on the interlayer As-As bond. For the meta-GGA calculations, we used the strongly constrained and appropriately normed (SCAN) functional \cite{sun2015} as this was constructed with consideration of $D(\mathbf{r})/D^{0}(\mathbf{r})$, on which the ELF is also dependent in Eq.~\ref{e:elf}.

\begin{table}[t]
\begin{tabular}{c|cc}
\hspace*{0.30in} & CT & TET    \\
\hline
\hline
 P (GPa) & 0.35 & 0.0   \\
 T (K) & 50 & 250   \\
 a (\AA) & 3.9792 & 3.8915   \\
 c (\AA) & 10.6379 & 11.690   \\
 z$_{\rm As}$ & 0.3687 & 0.372   \\
 Fe-As (\AA) & 2.3560 & 2.410   \\
 Fe-Fe (\AA) & 2.8137 & 2.7517   \\
\hline
\end{tabular}
\caption{\label{t:exp} The experimental structural parameters for the tetragonal (TET) and collapsed tetragonal (CT) structures from Ref. \onlinecite{Kreyssig2008}.}
\end{table}

Here, we considered the TET structure at 0 GPa and 250 K, and the CT structure at 0.35 GPa and 50K using the experimental structural parameters \cite{Kreyssig2008}, summarised in Table \ref{t:exp}. The TET and CT structures are shown in Figs.~\ref{f:CFA_DOS} (a) and (e), respectively, with both structures showing the parallelepiped Wigner-Seitz unit cell used in the calculations. Both the GGA and SCAN calculations used a 24$\times$24$\times$24 k-mesh for both structures. Whenever we discuss the GGA results, we are referring to the PBE functional. The FCSC DFT+DMFT calculations (using the PBE GGA functional) imposed DMFT on the Fe-d states using Wannier projectors generated within a correlated energy window around the Fermi energy of [-5.9, 16.0] ([-6.3, 16.0]) eV to encapsulate all of these Fe-d states in the TET (CT) structure. The DMFT calculations used the full rotationally invariant form of the interaction Hamiltonian with the CTQMC solver employing $2.52 \times 10^8$ sweeps. These DMFT calculations were in the paramagnetic phase and used the previously defined values of $U$~=~4.0~eV, $J$~=~0.8~eV, inverse temperature $\beta$~=~40~eV$^{-1}$, and the FLL double counting term \cite{diehl2014correlation}. Here, the FLL was calculated from the DMFT occupations.

\begin{figure}[h]
\centerline{\includegraphics[width=0.95\linewidth]{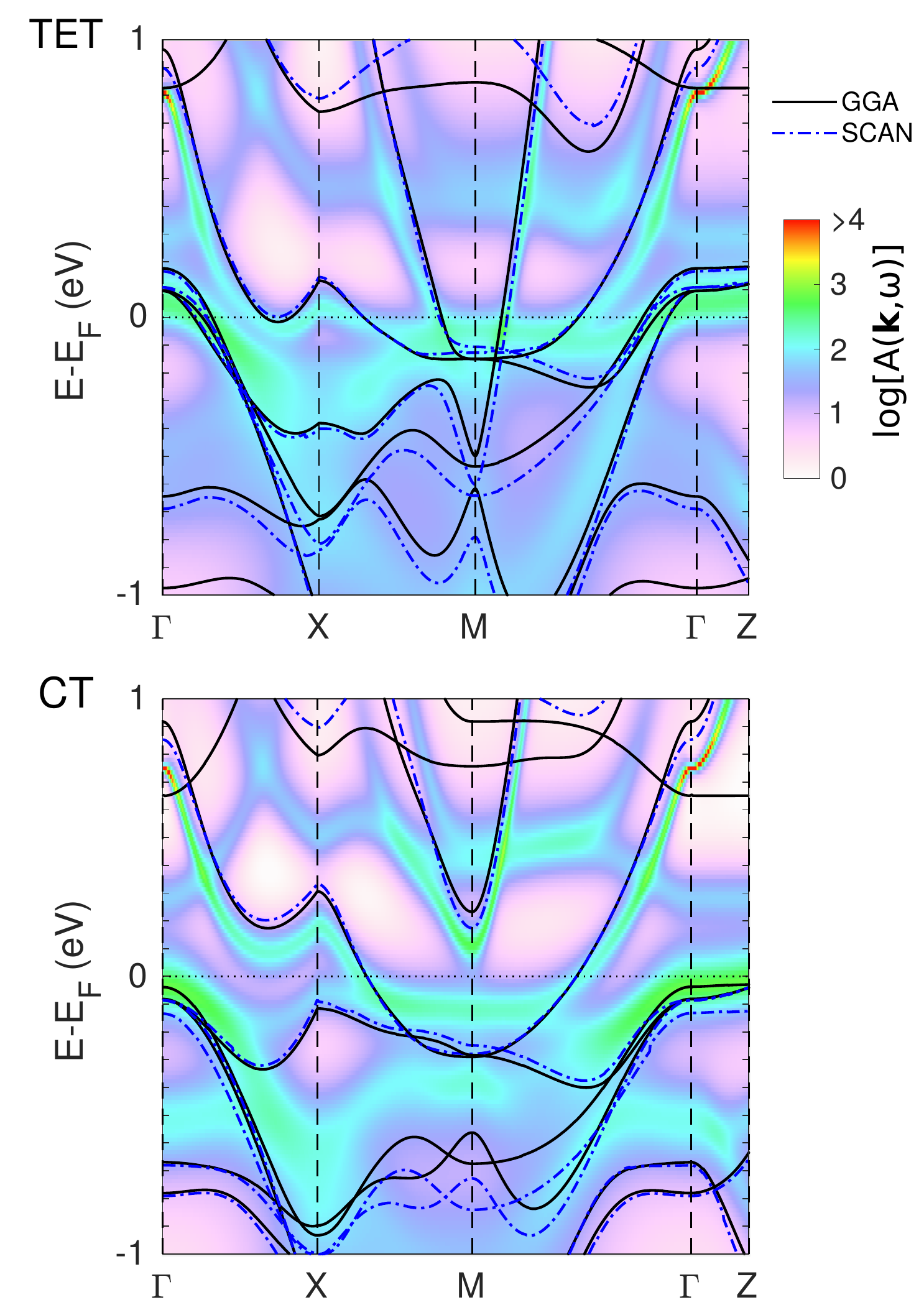}}
\caption{The bandstructures according to the GGA (PBE) and SCAN functionals and the $A(\textbf{k},\omega)$ of the fully charge self-consistent DFT+DMFT calculations for the tetragonal (TET) and collapsed tetragonal (CT) structures. The high symmetry points correspond to a simple tetragonal unit cell. The natural log color scale and range was used for clarity of the bands.}
\label{f:CFA_Akw}
\end{figure}

\begin{table}[t]
\begin{tabular}{c|cccc}
 &  & \hspace*{0.2in} $Z$ & & \\
\hspace*{0.30in}  & d$_{\rm z^2}$ & d$_{\rm x^2-y^2}$ & d$_{\rm xy}$ & d$_{\rm xz/yz}$   \\
\hline
\hline
TET & 0.57 & 0.53 & 0.60 & 0.54  \\
CT & 0.66 & 0.67 & 0.65 & 0.61  \\
\hline
\end{tabular}
\caption{\label{t:Z} The fully charge self-consistent DFT+DMFT quasiparticle residues ($Z$) in the tetragonal (TET) and collapsed tetragonal (CT) structures.}
\end{table}

Figs.~\ref{f:CFA_DOS} (b)/(f), (c)/(g) and (d)/(h) show the TET/CT total and partial As-p and Fe-d density of states (DOS) for GGA, SCAN, and FCSC DFT+DMFT calculations, respectively. The FCSC DOS was calculated using the analytically continued DMFT self-energy via the TRIQS/Maxent application \cite{maxent}. The SCAN DOS has a similar shape to the GGA, but the bandwidths have generally increased and the band centers tend to shift away from the Fermi level for larger absolute energies. This indicates that these states are more delocalized than in GGA functional calculations. On the other hand, the FCSC DOS shows a significant renormalization of the states around the Fermi level, as expected from the quasiparticle residue (see Table \ref{t:Z}) which broadly agrees with the previous studies on this material \cite{diehl2014correlation,Mandal2014}. The smooth profile is a consequence of the reduced quasiparticle lifetimes relating to the imaginary part of the self-energy. 

The bandstructures and $A(\textbf{k},\omega)$ for the TET and CT structures are shown in Figs.~\ref{f:CFA_Akw} (a) and (b), respectively. The FCSC $A(\textbf{k},\omega)$ broadly agrees with the previous studies in Refs. \onlinecite{vanroekeghem2016,Mandal2014}, and any discrepancies with those results are likely due to the different $U$, $J$, and correlated energy window used. Even within the displayed energy window around the Fermi level, the increased bandwidths of the SCAN bands can be seen. The band renormalization is distinct for the FCSC $A(\textbf{k},\omega)$. The GGA and SCAN bands cross the Fermi level at similar $\mathbf{k}$ values for the CT structure whereas there are small, but observable, changes for the TET structure. However, the FCSC $A(\textbf{k},\omega)$ have bands crossing the Fermi level at different $\mathbf{k}$ values compared to both the GGA and SCAN results for both structures.

\begin{figure}[h]
\centerline{\includegraphics[width=0.95\linewidth]{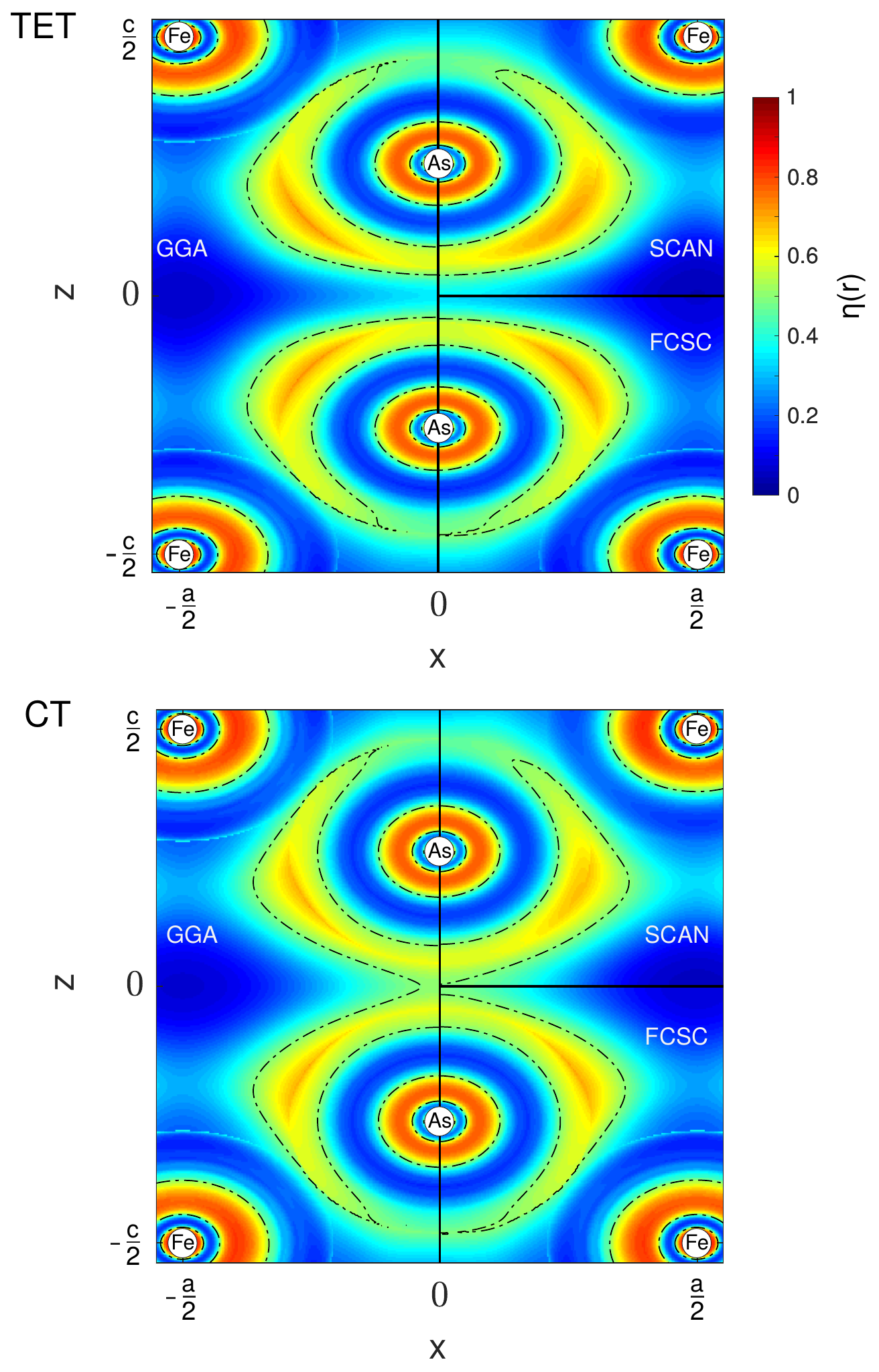}}
\caption{The 2D ELFs in the xz-plane (as indicated in Figs.~\ref{f:CFA_DOS} (a) and (e)) for the tetragonal (TET) and collapsed tetragonal (CT) structures. The ELFs for the GGA (PBE), SCAN and fully charge self-consistent (FCSC) DFT+DMFT calculations are shown in quadrants, as this plane is four-fold symmetric. The dashed contours are of $\eta(\textbf{r)}$~=~0.5.}
\label{f:CFA_ELFS}
\end{figure}

\begin{table}[t]
\begin{tabular}{cc|ccc}
\hspace*{0.30in} &  & GGA (PBE) & SCAN & FCSC   \\
\hline
\hline
TET & Fe-d (\%) & 25.2 & 23.4 & 31.8   \\
 & As-p (\%) & 26.2 & 26.8 & 23.0  \\
 & As-p limits (eV) & [-6.07, -1.93] & [-6.64, -2.15] & [-5.97, -2.03]  \\
 \hline
CT & Fe-d (\%)   & 24.5 & 22.5 & 29.7  \\
& As-p (\%)   & 25.7 & 26.5 & 23.4  \\
 & As-p limits (eV) & [-6.26, -1.99] & [-6.69, -2.15] & [-6.22, -2.33]  \\
\hline
\end{tabular}
\caption{\label{t:n_est} The estimated percentage of charge associated with As-p and Fe-d character for each calculation and structure. These percentages were calculated from the partial and total DOS integrals within the $\sigma$ bond DOS energy window. The As-p estimated energy ranges, which the DOS integrals were evaluated within, are also included.}
\end{table}

The ELFs for the TET and CT structures are given in Figs.~\ref{f:CFA_ELFS} (a) and (b), respectively. These ELFs are 2D slices in the xz-plane, which is indicated by the dashed lines in Figs.~\ref{f:CFA_DOS} (a) and (e), centered around the As-As bond. For the TET structure, the ELF indicates that there is a higher likelihood of finding an electron in the horseshoe shaped region between the interlayer As atoms. However, it is in the CT structure where these horseshoe shaped regions have coalesced to form distinct ELF weight at the center between the atoms. This indicates that the bond has formed between these interlayer As atoms. The ELFs in these regions are robust between the different calculations as significant (hybridized) As-p and Fe-d states (within the energy range of [-7, -2] eV) are still fully occupied meaning that the associated As-As and As-Fe bonds still exist. Although the $\mathbf{k}_{\rm F}$ of certain bands has changed between each calculation, this has not affected the ELF significantly, especially in the region around the Fe atoms as it is the states around the Fermi level which are predominantly Fe-d in character. However, there are slight changes to the ELF which are highlighted by the dot-dashed contour of $\eta(\textbf{r)}$~=~0.5. Compared to GGA, the SCAN $\eta(\textbf{r)}$ redistributes weight away from the Fe-As bond and moves it into the interstitial region away from Fe and As atoms. On the other hand, the FCSC $\eta(\textbf{r)}$ results redistribute weight (compared to GGA) to the region between the As and Fe atoms.  

These results are consistent with the changes in the percentages each atom contributes to the total charge within the interlayer As-As $\sigma$ bond regions (see Table \ref{t:n_est}). These were estimated by integrating the partial and total DOS over the energy range (see Table \ref{t:n_est}) in which there is significant As-p partial DOS. This energy range is where the (hybridized) As-p and Fe-d states are associated with the As-As and As-Fe bonds. Compared to the GGA, the SCAN results indicate that charge has redistributed away from the Fe atoms, which results in greater interstitial region contributions to the ELF. This is likely due to the delocalization of the states seen in the SCAN DOS and bandstructures. However, the FCSC results have greater charge around the Fe atom and reduced contributions from the other regions with respect to the GGA results. This increases the Fe-d and As-p hybridization and strengthens the Fe-As bond which in turn weakens the As-As bond. This is similar to the effect of the Fe spin-state in  Ref.~\onlinecite{Yildirim2009}, but here the strength of the bonds is weakly influenced by correlations. Both the SCAN and FCSC results have drawn weight away from the center of the As-As bond. Nonetheless, these calculations give an indication that the electron-electron correlation effects affect the As-p and Fe-d hybridization (as shown in Table \ref{t:n_est}), which in turn influences the strength of the As-As and Fe-As bonds as seen in the ELF.

\section{Conclusion}

We have introduced the newly developed {\sc Elk}-TRIQS interface which, in addition to standard one-particle quantities like spectral functions, also allows the calculation of orthonormal DFT+DMFT wavefunctions and related quantities. To illustrate the effect of correlations on these wavefunctions and occupations, we calculated the ELF.  The changes in the DFT+DMFT ELF come from the redistribution of the charge and kinetic energy density. The modifications in the charge density distribution dominated the changes in the ELF for monolayer SrVO$_3$, but it should be noted that in other materials this may not always be the case due to its dependence on the kinetic energy density. Therefore, a comparison of ELFs which have been computed within DFT+DMFT and DFT, as has been demonstrated in CaFe$_2$As$_2$, helps to visualize the role of correlations on the ELF. This means that the ELF, which is related to the strength of the chemical bonds, could be used to investigate changes in chemical bonding due to the electronic correlations from DMFT.

The {\sc Elk}-TRIQS monolayer SrVO$_3$ results have excellent agreement with the WIEN2k-TRIQS results and the generated ELFs were sensitive to the charge distribution of the V d$_{\rm xy}$ and d$_{\rm xz(yz)}$  orbitals. Therefore, any factors which influence the charge redistribution such as the correlated energy window and/or double counting, will also impact the DFT+DMFT ELF. The CaFe$_2$As$_2$ ELFs show an excellent visualization of the interlayer As-As bond formed in the CT phase. The Fe-As bonds (and As-As bonds present in the CT structure) are influenced by the Fe-d and As-p hybridization which is weakly dependent on the different electron-electron correlation effects beyond GGA that are present in either SCAN or FCSC DFT+DMFT calculations. This is contrary to the large effects that correlations have on the FCSC DFT+DMFT spectral functions. Correlations may have a more pronounced effect on the ELF if there is a significant charge redistribution and/or if the states affiliated with the bonds are at the Fermi level, where the additional correlations (from the many-body technique) affect the results more significantly. This could potentially be seen in other AB$_2$X$_2$ compounds, like those investigated in Ref \onlinecite{Cuervo-Reyes_2014}. We also suggest that the ELF should be used to investigate the effect of correlations on the bond-disproportionated insulating phase in rare-earth nickelates (RNiO$_3$) \cite{Peil_2019}. As well as this, the effect the MIT has on the bonds in heterostuctures, such as SrVO$_3$/SrTiO$_3$ \cite{zhong2015, james2020quantum}, and at their interfaces could be investigated using the ELF. Nonetheless, these results show the usefulness of the ELF in helping to visualize the bonds present in many other similarly correlated crystalline materials. 

There are other wavefunction dependent quantities which can be calculated using this formalism, such as the electron momentum density (EMD) which is experimentally probed by Compton scattering \cite{cooper2004x}. There is also the possibility of incorporating DMFT with other capabilities within {\sc Elk} such as the out-of-equilibrium functionality using TDDFT \cite{Sharma_2014}, where real material TDDFT+DMFT calculations are being reported \cite{Galicia_Hernandez_2020}.

\section{Acknowledgements}
A.~D.~N.~J.\ and E.~I.~H-L. acknowledge funding and support from the
Engineering and Physical Sciences Research Council (EPSRC). M.~A.\
acknowledges financial support from the Austrian Science Fund (FWF), START
program Y746. A.~H. acknowledges the Flatiron Institute which is a division of the Simons Foundation. The authors would like to acknowledge Dr.~M.~Zingl and the developers of TRIQS and Dr.~J.~K.~Dewhurst and the {\sc Elk} developers for their valuable contributions.
Calculations were performed using the computational facilities of the Advanced
Computing Research Centre, University of Bristol
(\href{http://bris.ac.uk/acrc/}{http://bris.ac.uk/acrc/}). The VESTA package
(\href{https://jp-minerals.org/vesta/en/}{https://jp-minerals.org/vesta/en/}) has been used in the preparation of some figures.

\appendix*
\section{Symmetries} \label{appendix:sym}
This section will briefly discuss how the symmetries have been used to construct the projectors and calculate the Brillouin zone integral of k-dependent quantities, $\mathbf{A}(\mathbf{k})$, in TRIQS from {\sc Elk}. 

For equivalent atoms in the unit cell, their second variational wavefunctions are equivalent. Therefore, the temporary projectors need to be transformed into the global basis before they are orthonormalized. This is done by using the matrix transformation of 
\begin{equation}
 \widetilde{P}_{m \nu}^{\text{global}, \alpha^{\prime}, \sigma}(\mathbf{k}) =  \mathcal{S}^{\alpha,  \alpha^{\prime}} \widetilde{P}_{m \nu}^{\text{loc}, \alpha, \sigma}(\mathbf{k}), 
\end{equation}
with $\mathcal{S}^{\alpha,  \alpha^{\prime}}$ specifying the symmetry matrix which transforms the projector to the equivalent atom site. The DMFT cycle will transform between the global and local coordinates. The global and local indices have been omitted in the main text for clarity. It should be noted that an $m$ orbital basis transformation will also be applied here (after the symmetry transformation). Therefore, a subset of $m$ values (such as the t$_{\rm 2g}$ orbitals) could be chosen and/or a diagonal $m$ basis could be used to help reduce issues with the sign problem (when using the CTQMC solver). 

The Brillouin zone integrals are calculated using the following generic formula
\begin{equation}
\sum_{\mathbf{k}}^{B Z} \mathbf{A}(\mathbf{k})=\sum_{s=1}^{N_{s}} \sum_{\mathbf{k}} \mathcal{S}_{s} \mathbf{A}(\mathbf{k}) \mathcal{S}_{s}^{\dagger},
\end{equation}
with $N_{s}$ symmetry operations.


\begin{thebibliography}{71}%
\makeatletter
\providecommand \@ifxundefined [1]{%
 \@ifx{#1\undefined}
}%
\providecommand \@ifnum [1]{%
 \ifnum #1\expandafter \@firstoftwo
 \else \expandafter \@secondoftwo
 \fi
}%
\providecommand \@ifx [1]{%
 \ifx #1\expandafter \@firstoftwo
 \else \expandafter \@secondoftwo
 \fi
}%
\providecommand \natexlab [1]{#1}%
\providecommand \enquote  [1]{``#1''}%
\providecommand \bibnamefont  [1]{#1}%
\providecommand \bibfnamefont [1]{#1}%
\providecommand \citenamefont [1]{#1}%
\providecommand \href@noop [0]{\@secondoftwo}%
\providecommand \href [0]{\begingroup \@sanitize@url \@href}%
\providecommand \@href[1]{\@@startlink{#1}\@@href}%
\providecommand \@@href[1]{\endgroup#1\@@endlink}%
\providecommand \@sanitize@url [0]{\catcode `\\12\catcode `\$12\catcode
  `\&12\catcode `\#12\catcode `\^12\catcode `\_12\catcode `\%12\relax}%
\providecommand \@@startlink[1]{}%
\providecommand \@@endlink[0]{}%
\providecommand \url  [0]{\begingroup\@sanitize@url \@url }%
\providecommand \@url [1]{\endgroup\@href {#1}{\urlprefix }}%
\providecommand \urlprefix  [0]{URL }%
\providecommand \Eprint [0]{\href }%
\providecommand \doibase [0]{http://dx.doi.org/}%
\providecommand \selectlanguage [0]{\@gobble}%
\providecommand \bibinfo  [0]{\@secondoftwo}%
\providecommand \bibfield  [0]{\@secondoftwo}%
\providecommand \translation [1]{[#1]}%
\providecommand \BibitemOpen [0]{}%
\providecommand \bibitemStop [0]{}%
\providecommand \bibitemNoStop [0]{.\EOS\space}%
\providecommand \EOS [0]{\spacefactor3000\relax}%
\providecommand \BibitemShut  [1]{\csname bibitem#1\endcsname}%
\let\auto@bib@innerbib\@empty
\bibitem [{\citenamefont {Perdew}\ and\ \citenamefont
  {Wang}(1992)}]{perdew1992}%
  \BibitemOpen
  \bibfield  {author} {\bibinfo {author} {\bibfnamefont {J.~P.}\ \bibnamefont
  {Perdew}}\ and\ \bibinfo {author} {\bibfnamefont {Y.}~\bibnamefont {Wang}},\
  }\bibfield  {title} {\enquote {\bibinfo {title} {Accurate and simple analytic
  representation of the electron-gas correlation energy},}\ }\href {\doibase
  10.1103/PhysRevB.45.13244} {\bibfield  {journal} {\bibinfo  {journal} {Phys.
  Rev. B}\ }\textbf {\bibinfo {volume} {45}},\ \bibinfo {pages} {13244--13249}
  (\bibinfo {year} {1992})}\BibitemShut {NoStop}%
\bibitem [{\citenamefont {Perdew}\ \emph
  {et~al.}(1996{\natexlab{a}})\citenamefont {Perdew}, \citenamefont {Burke},\
  and\ \citenamefont {Ernzerhof}}]{perdew1996}%
  \BibitemOpen
  \bibfield  {author} {\bibinfo {author} {\bibfnamefont {J.~P.}\ \bibnamefont
  {Perdew}}, \bibinfo {author} {\bibfnamefont {K.}~\bibnamefont {Burke}}, \
  and\ \bibinfo {author} {\bibfnamefont {M.}~\bibnamefont {Ernzerhof}},\
  }\bibfield  {title} {\enquote {\bibinfo {title} {Generalized gradient
  approximation made simple},}\ }\href
  {https://doi.org/10.1103/PhysRevLett.77.3865} {\bibfield  {journal} {\bibinfo
   {journal} {Phys. Rev. Lett.}\ }\textbf {\bibinfo {volume} {77}},\ \bibinfo
  {pages} {3865} (\bibinfo {year} {1996}{\natexlab{a}})}\BibitemShut {NoStop}%
\bibitem [{\citenamefont {Perdew}\ \emph
  {et~al.}(1996{\natexlab{b}})\citenamefont {Perdew}, \citenamefont {Burke},\
  and\ \citenamefont {Wang}}]{PBE1996}%
  \BibitemOpen
  \bibfield  {author} {\bibinfo {author} {\bibfnamefont {J.~P.}\ \bibnamefont
  {Perdew}}, \bibinfo {author} {\bibfnamefont {K}~\bibnamefont {Burke}}, \ and\
  \bibinfo {author} {\bibfnamefont {Y.}~\bibnamefont {Wang}},\ }\bibfield
  {title} {\enquote {\bibinfo {title} {Generalized gradient approximation for
  the exchange-correlation hole of a many-electron system},}\ }\href {\doibase
  10.1103/PhysRevB.54.16533} {\bibfield  {journal} {\bibinfo  {journal} {Phys.
  Rev. B}\ }\textbf {\bibinfo {volume} {54}},\ \bibinfo {pages} {16533--16539}
  (\bibinfo {year} {1996}{\natexlab{b}})}\BibitemShut {NoStop}%
\bibitem [{\citenamefont {Kurth}\ \emph {et~al.}(1999)\citenamefont {Kurth},
  \citenamefont {Perdew},\ and\ \citenamefont {Blaha}}]{kurth1999}%
  \BibitemOpen
  \bibfield  {author} {\bibinfo {author} {\bibfnamefont {S.}~\bibnamefont
  {Kurth}}, \bibinfo {author} {\bibfnamefont {J.~P.}\ \bibnamefont {Perdew}}, \
  and\ \bibinfo {author} {\bibfnamefont {P.}~\bibnamefont {Blaha}},\ }\bibfield
   {title} {\enquote {\bibinfo {title} {Molecular and solid-state tests of
  density functional approximations: {LSD}, {GGAs}, and {meta-GGAs}},}\ }\href
  {https://doi.org/10.1002/(SICI)1097-461X(1999)75:4/5<889::AID-QUA54>3.0.CO;2-8}
  {\bibfield  {journal} {\bibinfo  {journal} {International Journal of Quantum
  Chemistry}\ }\textbf {\bibinfo {volume} {75}},\ \bibinfo {pages} {889--909}
  (\bibinfo {year} {1999})}\BibitemShut {NoStop}%
\bibitem [{\citenamefont {Sun}\ \emph {et~al.}(2015)\citenamefont {Sun},
  \citenamefont {Ruzsinszky},\ and\ \citenamefont {Perdew}}]{sun2015}%
  \BibitemOpen
  \bibfield  {author} {\bibinfo {author} {\bibfnamefont {J.}~\bibnamefont
  {Sun}}, \bibinfo {author} {\bibfnamefont {A.}~\bibnamefont {Ruzsinszky}}, \
  and\ \bibinfo {author} {\bibfnamefont {J.~P.}\ \bibnamefont {Perdew}},\
  }\bibfield  {title} {\enquote {\bibinfo {title} {Strongly constrained and
  appropriately normed semilocal density functional},}\ }\href {\doibase
  10.1103/PhysRevLett.115.036402} {\bibfield  {journal} {\bibinfo  {journal}
  {Phys. Rev. Lett.}\ }\textbf {\bibinfo {volume} {115}},\ \bibinfo {pages}
  {036402} (\bibinfo {year} {2015})}\BibitemShut {NoStop}%
\bibitem [{\citenamefont {Perdew}\ \emph
  {et~al.}(1996{\natexlab{c}})\citenamefont {Perdew}, \citenamefont
  {Ernzerhof},\ and\ \citenamefont {Burke}}]{Perdew_1996_3}%
  \BibitemOpen
  \bibfield  {author} {\bibinfo {author} {\bibfnamefont {J.~P.}\ \bibnamefont
  {Perdew}}, \bibinfo {author} {\bibfnamefont {M.}~\bibnamefont {Ernzerhof}}, \
  and\ \bibinfo {author} {\bibfnamefont {K.}~\bibnamefont {Burke}},\ }\bibfield
   {title} {\enquote {\bibinfo {title} {Rationale for mixing exact exchange
  with density functional approximations},}\ }\href {\doibase 10.1063/1.472933}
  {\bibfield  {journal} {\bibinfo  {journal} {The Journal of Chemical Physics}\
  }\textbf {\bibinfo {volume} {105}},\ \bibinfo {pages} {9982--9985} (\bibinfo
  {year} {1996}{\natexlab{c}})}\BibitemShut {NoStop}%
\bibitem [{\citenamefont {Heyd}\ \emph {et~al.}(2003)\citenamefont {Heyd},
  \citenamefont {Scuseria},\ and\ \citenamefont {Ernzerhof}}]{Hyed_2003}%
  \BibitemOpen
  \bibfield  {author} {\bibinfo {author} {\bibfnamefont {J.}~\bibnamefont
  {Heyd}}, \bibinfo {author} {\bibfnamefont {G.~E.}\ \bibnamefont {Scuseria}},
  \ and\ \bibinfo {author} {\bibfnamefont {M.}~\bibnamefont {Ernzerhof}},\
  }\bibfield  {title} {\enquote {\bibinfo {title} {Hybrid functionals based on
  a screened coulomb potential},}\ }\href {\doibase 10.1063/1.1564060}
  {\bibfield  {journal} {\bibinfo  {journal} {The Journal of Chemical Physics}\
  }\textbf {\bibinfo {volume} {118}},\ \bibinfo {pages} {8207--8215} (\bibinfo
  {year} {2003})}\BibitemShut {NoStop}%
\bibitem [{\citenamefont {Anisimov}\ \emph {et~al.}(1997)\citenamefont
  {Anisimov}, \citenamefont {Poteryaev}, \citenamefont {Korotin}, \citenamefont
  {Anokhin},\ and\ \citenamefont {Kotliar}}]{Anisimov1997}%
  \BibitemOpen
  \bibfield  {author} {\bibinfo {author} {\bibfnamefont {V.~I.}\ \bibnamefont
  {Anisimov}}, \bibinfo {author} {\bibfnamefont {A.~I.}\ \bibnamefont
  {Poteryaev}}, \bibinfo {author} {\bibfnamefont {M.~A.}\ \bibnamefont
  {Korotin}}, \bibinfo {author} {\bibfnamefont {A.~O.}\ \bibnamefont
  {Anokhin}}, \ and\ \bibinfo {author} {\bibfnamefont {G.}~\bibnamefont
  {Kotliar}},\ }\bibfield  {title} {\enquote {\bibinfo {title}
  {First-principles calculations of the electronic structure and spectra of
  strongly correlated systems: dynamical mean-field theory},}\ }\href {\doibase
  10.1088/0953-8984/9/35/010} {\bibfield  {journal} {\bibinfo  {journal}
  {Journal of Physics: Condensed Matter}\ }\textbf {\bibinfo {volume} {9}},\
  \bibinfo {pages} {7359--7367} (\bibinfo {year} {1997})}\BibitemShut {NoStop}%
\bibitem [{\citenamefont {Lichtenstein}\ and\ \citenamefont
  {Katsnelson}(1998)}]{litchenstein1998}%
  \BibitemOpen
  \bibfield  {author} {\bibinfo {author} {\bibfnamefont {A.~I.}\ \bibnamefont
  {Lichtenstein}}\ and\ \bibinfo {author} {\bibfnamefont {M.~I.}\ \bibnamefont
  {Katsnelson}},\ }\bibfield  {title} {\enquote {\bibinfo {title} {Ab initio
  calculations of quasiparticle band structure in correlated systems: {LDA++}
  approach},}\ }\href {\doibase 10.1103/PhysRevB.57.6884} {\bibfield  {journal}
  {\bibinfo  {journal} {Phys. Rev. B}\ }\textbf {\bibinfo {volume} {57}},\
  \bibinfo {pages} {6884--6895} (\bibinfo {year} {1998})}\BibitemShut {NoStop}%
\bibitem [{\citenamefont {Aichhorn}\ \emph
  {et~al.}(2009{\natexlab{a}})\citenamefont {Aichhorn}, \citenamefont
  {Pourovskii}, \citenamefont {Vildosola}, \citenamefont {Ferrero},
  \citenamefont {Parcollet}, \citenamefont {Miyake}, \citenamefont {Georges},\
  and\ \citenamefont {Biermann}}]{aichhorn2009}%
  \BibitemOpen
  \bibfield  {author} {\bibinfo {author} {\bibfnamefont {M.}~\bibnamefont
  {Aichhorn}}, \bibinfo {author} {\bibfnamefont {L.}~\bibnamefont
  {Pourovskii}}, \bibinfo {author} {\bibfnamefont {V.}~\bibnamefont
  {Vildosola}}, \bibinfo {author} {\bibfnamefont {M.}~\bibnamefont {Ferrero}},
  \bibinfo {author} {\bibfnamefont {O.}~\bibnamefont {Parcollet}}, \bibinfo
  {author} {\bibfnamefont {T.}~\bibnamefont {Miyake}}, \bibinfo {author}
  {\bibfnamefont {A.}~\bibnamefont {Georges}}, \ and\ \bibinfo {author}
  {\bibfnamefont {S.}~\bibnamefont {Biermann}},\ }\bibfield  {title} {\enquote
  {\bibinfo {title} {Dynamical mean-field theory within an augmented plane-wave
  framework: Assessing electronic correlations in the iron pnictide
  {LaFeAsO}},}\ }\href {\doibase 10.1103/PhysRevB.80.085101} {\bibfield
  {journal} {\bibinfo  {journal} {Phys. Rev. B}\ }\textbf {\bibinfo {volume}
  {80}},\ \bibinfo {pages} {085101} (\bibinfo {year}
  {2009}{\natexlab{a}})}\BibitemShut {NoStop}%
\bibitem [{\citenamefont {Georges}\ \emph {et~al.}(1996)\citenamefont
  {Georges}, \citenamefont {Kotliar}, \citenamefont {Krauth},\ and\
  \citenamefont {Rozenberg}}]{georges1996}%
  \BibitemOpen
  \bibfield  {author} {\bibinfo {author} {\bibfnamefont {A.}~\bibnamefont
  {Georges}}, \bibinfo {author} {\bibfnamefont {G.}~\bibnamefont {Kotliar}},
  \bibinfo {author} {\bibfnamefont {W.}~\bibnamefont {Krauth}}, \ and\ \bibinfo
  {author} {\bibfnamefont {M.~J.}\ \bibnamefont {Rozenberg}},\ }\bibfield
  {title} {\enquote {\bibinfo {title} {Dynamical mean-field theory of strongly
  correlated fermion systems and the limit of infinite dimensions},}\ }\href
  {\doibase 10.1103/RevModPhys.68.13} {\bibfield  {journal} {\bibinfo
  {journal} {Rev. Mod. Phys.}\ }\textbf {\bibinfo {volume} {68}},\ \bibinfo
  {pages} {13--125} (\bibinfo {year} {1996})}\BibitemShut {NoStop}%
\bibitem [{\citenamefont {Becke}\ and\ \citenamefont
  {Edgecombe}(1990)}]{becke1990}%
  \BibitemOpen
  \bibfield  {author} {\bibinfo {author} {\bibfnamefont {A.~D.}\ \bibnamefont
  {Becke}}\ and\ \bibinfo {author} {\bibfnamefont {K.~E.}\ \bibnamefont
  {Edgecombe}},\ }\bibfield  {title} {\enquote {\bibinfo {title} {A simple
  measure of electron localization in atomic and molecular systems},}\ }\href
  {https://doi.org/10.1063/1.458517} {\bibfield  {journal} {\bibinfo  {journal}
  {The Journal of Chemical Physics}\ }\textbf {\bibinfo {volume} {92}},\
  \bibinfo {pages} {5397--5403} (\bibinfo {year} {1990})}\BibitemShut {NoStop}%
\bibitem [{\citenamefont {Savin}\ \emph {et~al.}(1997)\citenamefont {Savin},
  \citenamefont {Nesper}, \citenamefont {Wengert},\ and\ \citenamefont
  {Fässler}}]{Savin_1997}%
  \BibitemOpen
  \bibfield  {author} {\bibinfo {author} {\bibfnamefont {A.}~\bibnamefont
  {Savin}}, \bibinfo {author} {\bibfnamefont {R.}~\bibnamefont {Nesper}},
  \bibinfo {author} {\bibfnamefont {S.}~\bibnamefont {Wengert}}, \ and\
  \bibinfo {author} {\bibfnamefont {T.~F.}\ \bibnamefont {Fässler}},\
  }\bibfield  {title} {\enquote {\bibinfo {title} {{ELF}: The electron
  localization function},}\ }\href {\doibase 10.1002/anie.199718081} {\bibfield
   {journal} {\bibinfo  {journal} {Angewandte Chemie International Edition in
  English}\ }\textbf {\bibinfo {volume} {36}},\ \bibinfo {pages} {1808--1832}
  (\bibinfo {year} {1997})}\BibitemShut {NoStop}%
\bibitem [{\citenamefont {Silvi}\ \emph {et~al.}(2005)\citenamefont {Silvi},
  \citenamefont {Fourré},\ and\ \citenamefont {Alikhani}}]{Silvi2005}%
  \BibitemOpen
  \bibfield  {author} {\bibinfo {author} {\bibfnamefont {B.}~\bibnamefont
  {Silvi}}, \bibinfo {author} {\bibfnamefont {I.}~\bibnamefont {Fourré}}, \
  and\ \bibinfo {author} {\bibfnamefont {M.~E.}\ \bibnamefont {Alikhani}},\
  }\bibfield  {title} {\enquote {\bibinfo {title} {The topological analysis of
  the electron localization function. a key for a position space representation
  of chemical bonds},}\ }\href {\doibase 10.1007/s00706-005-0297-8} {\bibfield
  {journal} {\bibinfo  {journal} {Monatshefte für Chemie - Chemical Monthly}\
  }\textbf {\bibinfo {volume} {136}},\ \bibinfo {pages} {855--879} (\bibinfo
  {year} {2005})}\BibitemShut {NoStop}%
\bibitem [{\citenamefont {Santos}\ \emph {et~al.}(2005)\citenamefont {Santos},
  \citenamefont {Andres}, \citenamefont {Aizman},\ and\ \citenamefont
  {Fuentealba}}]{Santos_2005}%
  \BibitemOpen
  \bibfield  {author} {\bibinfo {author} {\bibfnamefont {J.~C.}\ \bibnamefont
  {Santos}}, \bibinfo {author} {\bibfnamefont {J.}~\bibnamefont {Andres}},
  \bibinfo {author} {\bibfnamefont {A.}~\bibnamefont {Aizman}}, \ and\ \bibinfo
  {author} {\bibfnamefont {P.}~\bibnamefont {Fuentealba}},\ }\bibfield  {title}
  {\enquote {\bibinfo {title} {An aromaticity scale based on the topological
  analysis of the electron localization function including {$\sigma$} and
  {$\pi$} contributions},}\ }\href {\doibase 10.1021/ct0499276} {\bibfield
  {journal} {\bibinfo  {journal} {Journal of Chemical Theory and Computation}\
  }\textbf {\bibinfo {volume} {1}},\ \bibinfo {pages} {83—86} (\bibinfo
  {year} {2005})}\BibitemShut {NoStop}%
\bibitem [{\citenamefont {Nalewajski}\ \emph {et~al.}(2005)\citenamefont
  {Nalewajski}, \citenamefont {Köster},\ and\ \citenamefont
  {Escalante}}]{Nalewajski_2005}%
  \BibitemOpen
  \bibfield  {author} {\bibinfo {author} {\bibfnamefont {R.~F.}\ \bibnamefont
  {Nalewajski}}, \bibinfo {author} {\bibfnamefont {A.~M.}\ \bibnamefont
  {Köster}}, \ and\ \bibinfo {author} {\bibfnamefont {S.}~\bibnamefont
  {Escalante}},\ }\bibfield  {title} {\enquote {\bibinfo {title} {Electron
  localization function as information measure},}\ }\href {\doibase
  10.1021/jp053184i} {\bibfield  {journal} {\bibinfo  {journal} {The Journal of
  Physical Chemistry. A}\ }\textbf {\bibinfo {volume} {109}},\ \bibinfo {pages}
  {10038—10043} (\bibinfo {year} {2005})}\BibitemShut {NoStop}%
\bibitem [{\citenamefont {Kohout}\ \emph {et~al.}(2002)\citenamefont {Kohout},
  \citenamefont {Wagner},\ and\ \citenamefont {Grin}}]{kohout2002electron}%
  \BibitemOpen
  \bibfield  {author} {\bibinfo {author} {\bibfnamefont {M.}~\bibnamefont
  {Kohout}}, \bibinfo {author} {\bibfnamefont {F.~R.}\ \bibnamefont {Wagner}},
  \ and\ \bibinfo {author} {\bibfnamefont {Y.}~\bibnamefont {Grin}},\
  }\bibfield  {title} {\enquote {\bibinfo {title} {Electron localization
  function for transition-metal compounds},}\ }\href
  {https://doi.org/10.1007/s00214-002-0370-x} {\bibfield  {journal} {\bibinfo
  {journal} {Theoretical Chemistry Accounts}\ }\textbf {\bibinfo {volume}
  {108}},\ \bibinfo {pages} {150--156} (\bibinfo {year} {2002})}\BibitemShut
  {NoStop}%
\bibitem [{\citenamefont {Cuervo-Reyes}\ and\ \citenamefont
  {Nesper}(2014)}]{Cuervo-Reyes_2014}%
  \BibitemOpen
  \bibfield  {author} {\bibinfo {author} {\bibfnamefont {E.}~\bibnamefont
  {Cuervo-Reyes}}\ and\ \bibinfo {author} {\bibfnamefont {R.}~\bibnamefont
  {Nesper}},\ }\bibfield  {title} {\enquote {\bibinfo {title} {Interlayer
  bonding in compounds with the {{ThCr}$_{2}${Si}$_{2}$}-type structure:
  Insight on the ferromagnetism of {{SrCo}$_{2}${({Ge}$_{\rm
  1\ensuremath{-}x}${P}$_{\rm x}$)}$_{2}$} from electronic structure
  calculations},}\ }\href {\doibase 10.1103/PhysRevB.90.064416} {\bibfield
  {journal} {\bibinfo  {journal} {Phys. Rev. B}\ }\textbf {\bibinfo {volume}
  {90}},\ \bibinfo {pages} {064416} (\bibinfo {year} {2014})}\BibitemShut
  {NoStop}%
\bibitem [{\citenamefont {Burnus}\ \emph {et~al.}(2005)\citenamefont {Burnus},
  \citenamefont {Marques},\ and\ \citenamefont {Gross}}]{burnus2005}%
  \BibitemOpen
  \bibfield  {author} {\bibinfo {author} {\bibfnamefont {T.}~\bibnamefont
  {Burnus}}, \bibinfo {author} {\bibfnamefont {M.~A.~L.}\ \bibnamefont
  {Marques}}, \ and\ \bibinfo {author} {\bibfnamefont {E.~K.~U.}\ \bibnamefont
  {Gross}},\ }\bibfield  {title} {\enquote {\bibinfo {title} {Time-dependent
  electron localization function},}\ }\href {\doibase
  10.1103/PhysRevA.71.010501} {\bibfield  {journal} {\bibinfo  {journal} {Phys.
  Rev. A}\ }\textbf {\bibinfo {volume} {71}},\ \bibinfo {pages} {010501(R)}
  (\bibinfo {year} {2005})}\BibitemShut {NoStop}%
\bibitem [{\citenamefont {Dewhurst}\ \emph {et~al.}()\citenamefont {Dewhurst},
  \citenamefont {Sharma}, \citenamefont {Nordstr\"{o}m}, \citenamefont
  {Cricchio}, \citenamefont {Granas},\ and\ \citenamefont {Gross}}]{elk}%
  \BibitemOpen
  \bibfield  {author} {\bibinfo {author} {\bibfnamefont {J.~K.}\ \bibnamefont
  {Dewhurst}}, \bibinfo {author} {\bibfnamefont {S.}~\bibnamefont {Sharma}},
  \bibinfo {author} {\bibfnamefont {L.}~\bibnamefont {Nordstr\"{o}m}}, \bibinfo
  {author} {\bibfnamefont {F.}~\bibnamefont {Cricchio}}, \bibinfo {author}
  {\bibfnamefont {O.}~\bibnamefont {Granas}}, \ and\ \bibinfo {author}
  {\bibfnamefont {E.~K.~U.}\ \bibnamefont {Gross}},\ }\href@noop {} {}\bibinfo
  {howpublished} {\url{http://elk.sourceforge.net/}}\BibitemShut {NoStop}%
\bibitem [{\citenamefont {Parcollet}\ \emph {et~al.}(2015)\citenamefont
  {Parcollet}, \citenamefont {Ferrero}, \citenamefont {Ayral}, \citenamefont
  {Hafermann}, \citenamefont {Krivenko}, \citenamefont {Messio},\ and\
  \citenamefont {Seth}}]{triqs}%
  \BibitemOpen
  \bibfield  {author} {\bibinfo {author} {\bibfnamefont {O.}~\bibnamefont
  {Parcollet}}, \bibinfo {author} {\bibfnamefont {M.}~\bibnamefont {Ferrero}},
  \bibinfo {author} {\bibfnamefont {T.}~\bibnamefont {Ayral}}, \bibinfo
  {author} {\bibfnamefont {H.}~\bibnamefont {Hafermann}}, \bibinfo {author}
  {\bibfnamefont {I.}~\bibnamefont {Krivenko}}, \bibinfo {author}
  {\bibfnamefont {L.}~\bibnamefont {Messio}}, \ and\ \bibinfo {author}
  {\bibfnamefont {P.}~\bibnamefont {Seth}},\ }\bibfield  {title} {\enquote
  {\bibinfo {title} {{TRIQS}: A toolbox for research on interacting quantum
  systems},}\ }\href {\doibase 10.1016/j.cpc.2015.04.023} {\bibfield  {journal}
  {\bibinfo  {journal} {Comp. Phys. Commun.}\ }\textbf {\bibinfo {volume}
  {196}},\ \bibinfo {pages} {398} (\bibinfo {year} {2015})}\BibitemShut
  {NoStop}%
\bibitem [{\citenamefont {Aichhorn}\ \emph {et~al.}()\citenamefont {Aichhorn},
  \citenamefont {Pourovskii}, \citenamefont {Seth}, \citenamefont {Vildosola},
  \citenamefont {Zingl}, \citenamefont {Peil}, \citenamefont {Deng},
  \citenamefont {Mravlje}, \citenamefont {Kraberger}, \citenamefont {Martins},
  \citenamefont {Ferrero},\ and\ \citenamefont {Parcollet}}]{DFTTools}%
  \BibitemOpen
  \bibfield  {author} {\bibinfo {author} {\bibfnamefont {M.}~\bibnamefont
  {Aichhorn}}, \bibinfo {author} {\bibfnamefont {L.}~\bibnamefont
  {Pourovskii}}, \bibinfo {author} {\bibfnamefont {P.}~\bibnamefont {Seth}},
  \bibinfo {author} {\bibfnamefont {V.}~\bibnamefont {Vildosola}}, \bibinfo
  {author} {\bibfnamefont {M.}~\bibnamefont {Zingl}}, \bibinfo {author}
  {\bibfnamefont {O.~E.}\ \bibnamefont {Peil}}, \bibinfo {author}
  {\bibfnamefont {X.}~\bibnamefont {Deng}}, \bibinfo {author} {\bibfnamefont
  {J.}~\bibnamefont {Mravlje}}, \bibinfo {author} {\bibfnamefont
  {G.}~\bibnamefont {Kraberger}}, \bibinfo {author} {\bibfnamefont
  {C.}~\bibnamefont {Martins}}, \bibinfo {author} {\bibfnamefont
  {M.}~\bibnamefont {Ferrero}}, \ and\ \bibinfo {author} {\bibfnamefont
  {O.}~\bibnamefont {Parcollet}},\ }\href@noop {} {}\bibinfo {howpublished}
  {\url{https://triqs.github.io/dft_tools}}\BibitemShut {NoStop}%
\bibitem [{\citenamefont {Aichhorn}\ \emph {et~al.}(2016)\citenamefont
  {Aichhorn}, \citenamefont {Pourovskii}, \citenamefont {Seth}, \citenamefont
  {Vildosola}, \citenamefont {Zingl}, \citenamefont {Peil}, \citenamefont
  {Deng}, \citenamefont {Mravlje}, \citenamefont {Kraberger}, \citenamefont
  {Martins}, \citenamefont {Ferrero},\ and\ \citenamefont
  {Parcollet}}]{aichhorn2016}%
  \BibitemOpen
  \bibfield  {author} {\bibinfo {author} {\bibfnamefont {M.}~\bibnamefont
  {Aichhorn}}, \bibinfo {author} {\bibfnamefont {L.}~\bibnamefont
  {Pourovskii}}, \bibinfo {author} {\bibfnamefont {P.}~\bibnamefont {Seth}},
  \bibinfo {author} {\bibfnamefont {V.}~\bibnamefont {Vildosola}}, \bibinfo
  {author} {\bibfnamefont {M.}~\bibnamefont {Zingl}}, \bibinfo {author}
  {\bibfnamefont {O.~E.}\ \bibnamefont {Peil}}, \bibinfo {author}
  {\bibfnamefont {X.}~\bibnamefont {Deng}}, \bibinfo {author} {\bibfnamefont
  {J.}~\bibnamefont {Mravlje}}, \bibinfo {author} {\bibfnamefont {G.~J.}\
  \bibnamefont {Kraberger}}, \bibinfo {author} {\bibfnamefont {C.}~\bibnamefont
  {Martins}}, \bibinfo {author} {\bibfnamefont {M.}~\bibnamefont {Ferrero}}, \
  and\ \bibinfo {author} {\bibfnamefont {O.}~\bibnamefont {Parcollet}},\
  }\bibfield  {title} {\enquote {\bibinfo {title} {{TRIQS/DFTTools}: A {TRIQS}
  application for ab initio calculations of correlated materials},}\ }\href
  {\doibase 10.1016/j.cpc.2016.03.014} {\bibfield  {journal} {\bibinfo
  {journal} {Comp. Phys. Commun.}\ }\textbf {\bibinfo {volume} {204}},\
  \bibinfo {pages} {200} (\bibinfo {year} {2016})}\BibitemShut {NoStop}%
\bibitem [{\citenamefont {Lechermann}\ \emph {et~al.}(2006)\citenamefont
  {Lechermann}, \citenamefont {Georges}, \citenamefont {Poteryaev},
  \citenamefont {Biermann}, \citenamefont {Posternak}, \citenamefont
  {Yamasaki},\ and\ \citenamefont {Andersen}}]{Lechermann2006}%
  \BibitemOpen
  \bibfield  {author} {\bibinfo {author} {\bibfnamefont {F.}~\bibnamefont
  {Lechermann}}, \bibinfo {author} {\bibfnamefont {A.}~\bibnamefont {Georges}},
  \bibinfo {author} {\bibfnamefont {A.}~\bibnamefont {Poteryaev}}, \bibinfo
  {author} {\bibfnamefont {S.}~\bibnamefont {Biermann}}, \bibinfo {author}
  {\bibfnamefont {M.}~\bibnamefont {Posternak}}, \bibinfo {author}
  {\bibfnamefont {A.}~\bibnamefont {Yamasaki}}, \ and\ \bibinfo {author}
  {\bibfnamefont {O.~K.}\ \bibnamefont {Andersen}},\ }\bibfield  {title}
  {\enquote {\bibinfo {title} {Dynamical mean-field theory using {Wannier}
  functions: A flexible route to electronic structure calculations of strongly
  correlated materials},}\ }\href {\doibase 10.1103/PhysRevB.74.125120}
  {\bibfield  {journal} {\bibinfo  {journal} {Phys. Rev. B}\ }\textbf {\bibinfo
  {volume} {74}},\ \bibinfo {pages} {125120} (\bibinfo {year}
  {2006})}\BibitemShut {NoStop}%
\bibitem [{\citenamefont {Amadon}\ \emph {et~al.}(2008)\citenamefont {Amadon},
  \citenamefont {Lechermann}, \citenamefont {Georges}, \citenamefont {Jollet},
  \citenamefont {Wehling},\ and\ \citenamefont {Lichtenstein}}]{amadon2008}%
  \BibitemOpen
  \bibfield  {author} {\bibinfo {author} {\bibfnamefont {B.}~\bibnamefont
  {Amadon}}, \bibinfo {author} {\bibfnamefont {F.}~\bibnamefont {Lechermann}},
  \bibinfo {author} {\bibfnamefont {A.}~\bibnamefont {Georges}}, \bibinfo
  {author} {\bibfnamefont {F.}~\bibnamefont {Jollet}}, \bibinfo {author}
  {\bibfnamefont {T.~O.}\ \bibnamefont {Wehling}}, \ and\ \bibinfo {author}
  {\bibfnamefont {A.~I.}\ \bibnamefont {Lichtenstein}},\ }\bibfield  {title}
  {\enquote {\bibinfo {title} {Plane-wave based electronic structure
  calculations for correlated materials using dynamical mean-field theory and
  projected local orbitals},}\ }\href {\doibase 10.1103/PhysRevB.77.205112}
  {\bibfield  {journal} {\bibinfo  {journal} {Phys. Rev. B}\ }\textbf {\bibinfo
  {volume} {77}},\ \bibinfo {pages} {205112} (\bibinfo {year}
  {2008})}\BibitemShut {NoStop}%
\bibitem [{\citenamefont {Aichhorn}\ \emph
  {et~al.}(2009{\natexlab{b}})\citenamefont {Aichhorn}, \citenamefont
  {Pourovskii}, \citenamefont {Vildosola}, \citenamefont {Ferrero},
  \citenamefont {Parcollet}, \citenamefont {Miyake}, \citenamefont {Georges},\
  and\ \citenamefont {Biermann}}]{aichhorn2009dynamical}%
  \BibitemOpen
  \bibfield  {author} {\bibinfo {author} {\bibfnamefont {M}~\bibnamefont
  {Aichhorn}}, \bibinfo {author} {\bibfnamefont {L}~\bibnamefont {Pourovskii}},
  \bibinfo {author} {\bibfnamefont {V}~\bibnamefont {Vildosola}}, \bibinfo
  {author} {\bibfnamefont {M}~\bibnamefont {Ferrero}}, \bibinfo {author}
  {\bibfnamefont {O}~\bibnamefont {Parcollet}}, \bibinfo {author}
  {\bibfnamefont {T}~\bibnamefont {Miyake}}, \bibinfo {author} {\bibfnamefont
  {A}~\bibnamefont {Georges}}, \ and\ \bibinfo {author} {\bibfnamefont
  {S}~\bibnamefont {Biermann}},\ }\bibfield  {title} {\enquote {\bibinfo
  {title} {Dynamical mean-field theory within an augmented plane-wave
  framework: Assessing electronic correlations in the iron pnictide
  {LaFeAsO}},}\ }\href {\doibase 10.1103/PhysRevB.80.085101} {\bibfield
  {journal} {\bibinfo  {journal} {Phys. Rev. B}\ }\textbf {\bibinfo {volume}
  {80}},\ \bibinfo {pages} {085101} (\bibinfo {year}
  {2009}{\natexlab{b}})}\BibitemShut {NoStop}%
\bibitem [{\citenamefont {Kurz}\ \emph {et~al.}(2004)\citenamefont {Kurz},
  \citenamefont {F\"orster}, \citenamefont {Nordstr\"om}, \citenamefont
  {Bihlmayer},\ and\ \citenamefont {Bl\"ugel}}]{Kurz_2004}%
  \BibitemOpen
  \bibfield  {author} {\bibinfo {author} {\bibfnamefont {Ph.}\ \bibnamefont
  {Kurz}}, \bibinfo {author} {\bibfnamefont {F.}~\bibnamefont {F\"orster}},
  \bibinfo {author} {\bibfnamefont {L.}~\bibnamefont {Nordstr\"om}}, \bibinfo
  {author} {\bibfnamefont {G.}~\bibnamefont {Bihlmayer}}, \ and\ \bibinfo
  {author} {\bibfnamefont {S.}~\bibnamefont {Bl\"ugel}},\ }\bibfield  {title}
  {\enquote {\bibinfo {title} {Ab initio treatment of noncollinear magnets with
  the full-potential linearized augmented plane wave method},}\ }\href
  {\doibase 10.1103/PhysRevB.69.024415} {\bibfield  {journal} {\bibinfo
  {journal} {Phys. Rev. B}\ }\textbf {\bibinfo {volume} {69}},\ \bibinfo
  {pages} {024415} (\bibinfo {year} {2004})}\BibitemShut {NoStop}%
\bibitem [{\citenamefont {Sch{\"u}ler}\ \emph {et~al.}(2018)\citenamefont
  {Sch{\"u}ler}, \citenamefont {Peil}, \citenamefont {Kraberger}, \citenamefont
  {Pordzik}, \citenamefont {Marsman}, \citenamefont {Kresse}, \citenamefont
  {Wehling},\ and\ \citenamefont {Aichhorn}}]{schuler2018}%
  \BibitemOpen
  \bibfield  {author} {\bibinfo {author} {\bibfnamefont {M.}~\bibnamefont
  {Sch{\"u}ler}}, \bibinfo {author} {\bibfnamefont {O.~E.}\ \bibnamefont
  {Peil}}, \bibinfo {author} {\bibfnamefont {G.~J.}\ \bibnamefont {Kraberger}},
  \bibinfo {author} {\bibfnamefont {R.}~\bibnamefont {Pordzik}}, \bibinfo
  {author} {\bibfnamefont {M.}~\bibnamefont {Marsman}}, \bibinfo {author}
  {\bibfnamefont {G.}~\bibnamefont {Kresse}}, \bibinfo {author} {\bibfnamefont
  {T.~O.}\ \bibnamefont {Wehling}}, \ and\ \bibinfo {author} {\bibfnamefont
  {M.}~\bibnamefont {Aichhorn}},\ }\bibfield  {title} {\enquote {\bibinfo
  {title} {Charge self-consistent many-body corrections using optimized
  projected localized orbitals},}\ }\href {\doibase 10.1088/1361-648X/aae80a}
  {\bibfield  {journal} {\bibinfo  {journal} {J. Phys.: Conden. Matter}\
  }\textbf {\bibinfo {volume} {30}},\ \bibinfo {pages} {475901} (\bibinfo
  {year} {2018})}\BibitemShut {NoStop}%
\bibitem [{\citenamefont {Savin}\ \emph {et~al.}(1992)\citenamefont {Savin},
  \citenamefont {Jepsen}, \citenamefont {Flad}, \citenamefont {Andersen},
  \citenamefont {Preuss},\ and\ \citenamefont {von Schnering}}]{Savin_1992}%
  \BibitemOpen
  \bibfield  {author} {\bibinfo {author} {\bibfnamefont {A.}~\bibnamefont
  {Savin}}, \bibinfo {author} {\bibfnamefont {O.}~\bibnamefont {Jepsen}},
  \bibinfo {author} {\bibfnamefont {J.}~\bibnamefont {Flad}}, \bibinfo {author}
  {\bibfnamefont {O.~K.}\ \bibnamefont {Andersen}}, \bibinfo {author}
  {\bibfnamefont {H.}~\bibnamefont {Preuss}}, \ and\ \bibinfo {author}
  {\bibfnamefont {H.~G.}\ \bibnamefont {von Schnering}},\ }\bibfield  {title}
  {\enquote {\bibinfo {title} {Electron localization in solid-state structures
  of the elements: the diamond structure},}\ }\href {\doibase
  10.1002/anie.199201871} {\bibfield  {journal} {\bibinfo  {journal}
  {Angewandte Chemie International Edition in English}\ }\textbf {\bibinfo
  {volume} {31}},\ \bibinfo {pages} {187--188} (\bibinfo {year}
  {1992})}\BibitemShut {NoStop}%
\bibitem [{\citenamefont {Kohout}\ and\ \citenamefont
  {Savin}(1997)}]{kohout1997influence}%
  \BibitemOpen
  \bibfield  {author} {\bibinfo {author} {\bibfnamefont {M.}~\bibnamefont
  {Kohout}}\ and\ \bibinfo {author} {\bibfnamefont {A.}~\bibnamefont {Savin}},\
  }\bibfield  {title} {\enquote {\bibinfo {title} {Influence of core--valence
  separation of electron localization function},}\ }\href
  {https://doi.org/10.1002/(SICI)1096-987X(199709)18:12<1431::AID-JCC1>3.0.CO;2-K}
  {\bibfield  {journal} {\bibinfo  {journal} {Journal of Computational
  Chemistry}\ }\textbf {\bibinfo {volume} {18}},\ \bibinfo {pages} {1431--1439}
  (\bibinfo {year} {1997})}\BibitemShut {NoStop}%
\bibitem [{\citenamefont {Nekrasov}\ \emph {et~al.}(2005)\citenamefont
  {Nekrasov}, \citenamefont {Keller}, \citenamefont {Kondakov}, \citenamefont
  {Kozhevnikov}, \citenamefont {Pruschke}, \citenamefont {Held}, \citenamefont
  {Vollhardt},\ and\ \citenamefont {Anisimov}}]{nekrasov2005}%
  \BibitemOpen
  \bibfield  {author} {\bibinfo {author} {\bibfnamefont {I.~A.}\ \bibnamefont
  {Nekrasov}}, \bibinfo {author} {\bibfnamefont {G.}~\bibnamefont {Keller}},
  \bibinfo {author} {\bibfnamefont {D.~E.}\ \bibnamefont {Kondakov}}, \bibinfo
  {author} {\bibfnamefont {A.~V.}\ \bibnamefont {Kozhevnikov}}, \bibinfo
  {author} {\bibfnamefont {Th.}\ \bibnamefont {Pruschke}}, \bibinfo {author}
  {\bibfnamefont {K.}~\bibnamefont {Held}}, \bibinfo {author} {\bibfnamefont
  {D.}~\bibnamefont {Vollhardt}}, \ and\ \bibinfo {author} {\bibfnamefont
  {V.~I.}\ \bibnamefont {Anisimov}},\ }\bibfield  {title} {\enquote {\bibinfo
  {title} {Comparative study of correlation effects in {CaVO}$_3$ and
  {SrVO}$_3$},}\ }\href {\doibase 10.1103/PhysRevB.72.155106} {\bibfield
  {journal} {\bibinfo  {journal} {Phys. Rev. B}\ }\textbf {\bibinfo {volume}
  {72}},\ \bibinfo {pages} {155106} (\bibinfo {year} {2005})}\BibitemShut
  {NoStop}%
\bibitem [{\citenamefont {Sekiyama}\ \emph {et~al.}(2004)\citenamefont
  {Sekiyama}, \citenamefont {Fujiwara}, \citenamefont {Imada}, \citenamefont
  {Suga}, \citenamefont {Eisaki}, \citenamefont {Uchida}, \citenamefont
  {Takegahara}, \citenamefont {Harima}, \citenamefont {Saitoh}, \citenamefont
  {Nekrasov}, \citenamefont {Keller}, \citenamefont {Kondakov}, \citenamefont
  {Kozhevnikov}, \citenamefont {Pruschke}, \citenamefont {Held}, \citenamefont
  {Vollhardt},\ and\ \citenamefont {Anisimov}}]{sekiyama2004}%
  \BibitemOpen
  \bibfield  {author} {\bibinfo {author} {\bibfnamefont {A.}~\bibnamefont
  {Sekiyama}}, \bibinfo {author} {\bibfnamefont {H.}~\bibnamefont {Fujiwara}},
  \bibinfo {author} {\bibfnamefont {S.}~\bibnamefont {Imada}}, \bibinfo
  {author} {\bibfnamefont {S.}~\bibnamefont {Suga}}, \bibinfo {author}
  {\bibfnamefont {H.}~\bibnamefont {Eisaki}}, \bibinfo {author} {\bibfnamefont
  {S.~I.}\ \bibnamefont {Uchida}}, \bibinfo {author} {\bibfnamefont
  {K.}~\bibnamefont {Takegahara}}, \bibinfo {author} {\bibfnamefont
  {H.}~\bibnamefont {Harima}}, \bibinfo {author} {\bibfnamefont
  {Y.}~\bibnamefont {Saitoh}}, \bibinfo {author} {\bibfnamefont {I.~A.}\
  \bibnamefont {Nekrasov}}, \bibinfo {author} {\bibfnamefont {G.}~\bibnamefont
  {Keller}}, \bibinfo {author} {\bibfnamefont {D.~E.}\ \bibnamefont
  {Kondakov}}, \bibinfo {author} {\bibfnamefont {A.~V.}\ \bibnamefont
  {Kozhevnikov}}, \bibinfo {author} {\bibfnamefont {Th.}\ \bibnamefont
  {Pruschke}}, \bibinfo {author} {\bibfnamefont {K.}~\bibnamefont {Held}},
  \bibinfo {author} {\bibfnamefont {D.}~\bibnamefont {Vollhardt}}, \ and\
  \bibinfo {author} {\bibfnamefont {V.~I.}\ \bibnamefont {Anisimov}},\
  }\bibfield  {title} {\enquote {\bibinfo {title} {Mutual experimental and
  theoretical validation of bulk photoemission spectra of
  {Sr}$_{1-x}${Ca}$_x${VO}$_3$},}\ }\href {\doibase
  10.1103/PhysRevLett.93.156402} {\bibfield  {journal} {\bibinfo  {journal}
  {Phys. Rev. Lett.}\ }\textbf {\bibinfo {volume} {93}},\ \bibinfo {pages}
  {156402} (\bibinfo {year} {2004})}\BibitemShut {NoStop}%
\bibitem [{\citenamefont {Nekrasov}\ \emph {et~al.}(2006)\citenamefont
  {Nekrasov}, \citenamefont {Held}, \citenamefont {Keller}, \citenamefont
  {Kondakov}, \citenamefont {Pruschke}, \citenamefont {Kollar}, \citenamefont
  {Andersen}, \citenamefont {Anisimov},\ and\ \citenamefont
  {Vollhardt}}]{nekrasov2006}%
  \BibitemOpen
  \bibfield  {author} {\bibinfo {author} {\bibfnamefont {I.~A.}\ \bibnamefont
  {Nekrasov}}, \bibinfo {author} {\bibfnamefont {K.}~\bibnamefont {Held}},
  \bibinfo {author} {\bibfnamefont {G.}~\bibnamefont {Keller}}, \bibinfo
  {author} {\bibfnamefont {D.~E.}\ \bibnamefont {Kondakov}}, \bibinfo {author}
  {\bibfnamefont {Th.}\ \bibnamefont {Pruschke}}, \bibinfo {author}
  {\bibfnamefont {M.}~\bibnamefont {Kollar}}, \bibinfo {author} {\bibfnamefont
  {O.~K.}\ \bibnamefont {Andersen}}, \bibinfo {author} {\bibfnamefont {V.~I.}\
  \bibnamefont {Anisimov}}, \ and\ \bibinfo {author} {\bibfnamefont
  {D.}~\bibnamefont {Vollhardt}},\ }\bibfield  {title} {\enquote {\bibinfo
  {title} {Momentum-resolved spectral functions of {SrVO}$_3$ calculated by
  {LDA+DMFT}},}\ }\href {\doibase 10.1103/PhysRevB.73.155112} {\bibfield
  {journal} {\bibinfo  {journal} {Phys. Rev. B}\ }\textbf {\bibinfo {volume}
  {73}},\ \bibinfo {pages} {155112} (\bibinfo {year} {2006})}\BibitemShut
  {NoStop}%
\bibitem [{\citenamefont {Kotliar}\ \emph {et~al.}(2006)\citenamefont
  {Kotliar}, \citenamefont {Savrasov}, \citenamefont {Haule}, \citenamefont
  {Oudovenko}, \citenamefont {Parcollet},\ and\ \citenamefont
  {Marianetti}}]{kotliar2006}%
  \BibitemOpen
  \bibfield  {author} {\bibinfo {author} {\bibfnamefont {G.}~\bibnamefont
  {Kotliar}}, \bibinfo {author} {\bibfnamefont {S.~Y.}\ \bibnamefont
  {Savrasov}}, \bibinfo {author} {\bibfnamefont {K.}~\bibnamefont {Haule}},
  \bibinfo {author} {\bibfnamefont {V.~S.}\ \bibnamefont {Oudovenko}}, \bibinfo
  {author} {\bibfnamefont {O.}~\bibnamefont {Parcollet}}, \ and\ \bibinfo
  {author} {\bibfnamefont {C.~A.}\ \bibnamefont {Marianetti}},\ }\bibfield
  {title} {\enquote {\bibinfo {title} {Electronic structure calculations with
  dynamical mean-field theory},}\ }\href
  {https://doi.org/10.1103/RevModPhys.78.865} {\bibfield  {journal} {\bibinfo
  {journal} {Rev. Mod. Phys.}\ }\textbf {\bibinfo {volume} {78}},\ \bibinfo
  {pages} {865} (\bibinfo {year} {2006})}\BibitemShut {NoStop}%
\bibitem [{\citenamefont {Byczuk}\ \emph {et~al.}(2007)\citenamefont {Byczuk},
  \citenamefont {Kollar}, \citenamefont {Held}, \citenamefont {Yang},
  \citenamefont {Nekrasov}, \citenamefont {Pruschke},\ and\ \citenamefont
  {Vollhardt}}]{byczuk2007}%
  \BibitemOpen
  \bibfield  {author} {\bibinfo {author} {\bibfnamefont {K.}~\bibnamefont
  {Byczuk}}, \bibinfo {author} {\bibfnamefont {M.}~\bibnamefont {Kollar}},
  \bibinfo {author} {\bibfnamefont {K.}~\bibnamefont {Held}}, \bibinfo {author}
  {\bibfnamefont {Y.-F.}\ \bibnamefont {Yang}}, \bibinfo {author}
  {\bibfnamefont {I.~A.}\ \bibnamefont {Nekrasov}}, \bibinfo {author}
  {\bibfnamefont {Th.}\ \bibnamefont {Pruschke}}, \ and\ \bibinfo {author}
  {\bibfnamefont {D.}~\bibnamefont {Vollhardt}},\ }\bibfield  {title} {\enquote
  {\bibinfo {title} {Kinks in the dispersion of strongly correlated
  electrons},}\ }\href {https://doi.org/10.1038/nphys538} {\bibfield  {journal}
  {\bibinfo  {journal} {Nat. Phys.}\ }\textbf {\bibinfo {volume} {3}},\
  \bibinfo {pages} {168} (\bibinfo {year} {2007})}\BibitemShut {NoStop}%
\bibitem [{\citenamefont {Tomczak}\ \emph {et~al.}(2012)\citenamefont
  {Tomczak}, \citenamefont {Casula}, \citenamefont {Miyake}, \citenamefont
  {Aryasetiawan},\ and\ \citenamefont {Biermann}}]{tomczak2012}%
  \BibitemOpen
  \bibfield  {author} {\bibinfo {author} {\bibfnamefont {J.~M.}\ \bibnamefont
  {Tomczak}}, \bibinfo {author} {\bibfnamefont {M.}~\bibnamefont {Casula}},
  \bibinfo {author} {\bibfnamefont {T.}~\bibnamefont {Miyake}}, \bibinfo
  {author} {\bibfnamefont {F.}~\bibnamefont {Aryasetiawan}}, \ and\ \bibinfo
  {author} {\bibfnamefont {S.}~\bibnamefont {Biermann}},\ }\bibfield  {title}
  {\enquote {\bibinfo {title} {Combined {GW} and dynamical mean-field theory:
  {D}ynamical screening effects in transition metal oxides},}\ }\href
  {https://doi.org/10.1209/0295-5075/100/67001} {\bibfield  {journal} {\bibinfo
   {journal} {Europhys. Lett.}\ }\textbf {\bibinfo {volume} {100}},\ \bibinfo
  {pages} {67001} (\bibinfo {year} {2012})}\BibitemShut {NoStop}%
\bibitem [{\citenamefont {Boehnke}\ \emph {et~al.}(2016)\citenamefont
  {Boehnke}, \citenamefont {Nilsson}, \citenamefont {Aryasetiawan},\ and\
  \citenamefont {Werner}}]{Boehnke2016}%
  \BibitemOpen
  \bibfield  {author} {\bibinfo {author} {\bibfnamefont {L.}~\bibnamefont
  {Boehnke}}, \bibinfo {author} {\bibfnamefont {F.}~\bibnamefont {Nilsson}},
  \bibinfo {author} {\bibfnamefont {F.}~\bibnamefont {Aryasetiawan}}, \ and\
  \bibinfo {author} {\bibfnamefont {P.}~\bibnamefont {Werner}},\ }\bibfield
  {title} {\enquote {\bibinfo {title} {When strong correlations become weak:
  Consistent merging of {$GW$} and {DMFT}},}\ }\href {\doibase
  10.1103/PhysRevB.94.201106} {\bibfield  {journal} {\bibinfo  {journal} {Phys.
  Rev. B}\ }\textbf {\bibinfo {volume} {94}},\ \bibinfo {pages} {201106(R)}
  (\bibinfo {year} {2016})}\BibitemShut {NoStop}%
\bibitem [{\citenamefont {Yoshimatsu}\ \emph {et~al.}(2010)\citenamefont
  {Yoshimatsu}, \citenamefont {Okabe}, \citenamefont {Kumigashira},
  \citenamefont {Okamoto}, \citenamefont {Aizaki}, \citenamefont {Fujimori},\
  and\ \citenamefont {Oshima}}]{yoshimatsu2010}%
  \BibitemOpen
  \bibfield  {author} {\bibinfo {author} {\bibfnamefont {K.}~\bibnamefont
  {Yoshimatsu}}, \bibinfo {author} {\bibfnamefont {T.}~\bibnamefont {Okabe}},
  \bibinfo {author} {\bibfnamefont {H.}~\bibnamefont {Kumigashira}}, \bibinfo
  {author} {\bibfnamefont {S.}~\bibnamefont {Okamoto}}, \bibinfo {author}
  {\bibfnamefont {S.}~\bibnamefont {Aizaki}}, \bibinfo {author} {\bibfnamefont
  {A.}~\bibnamefont {Fujimori}}, \ and\ \bibinfo {author} {\bibfnamefont
  {M.}~\bibnamefont {Oshima}},\ }\bibfield  {title} {\enquote {\bibinfo {title}
  {Dimensional-crossover-driven metal-insulator transition in {SrVO}$_3$
  ultrathin films},}\ }\href {\doibase 10.1103/PhysRevLett.104.147601}
  {\bibfield  {journal} {\bibinfo  {journal} {Phys. Rev. Lett.}\ }\textbf
  {\bibinfo {volume} {104}},\ \bibinfo {pages} {147601} (\bibinfo {year}
  {2010})}\BibitemShut {NoStop}%
\bibitem [{\citenamefont {Gu}\ \emph {et~al.}(2014)\citenamefont {Gu},
  \citenamefont {Wolf},\ and\ \citenamefont {Lu}}]{gu2014}%
  \BibitemOpen
  \bibfield  {author} {\bibinfo {author} {\bibfnamefont {M.}~\bibnamefont
  {Gu}}, \bibinfo {author} {\bibfnamefont {S.~A.}\ \bibnamefont {Wolf}}, \ and\
  \bibinfo {author} {\bibfnamefont {J.}~\bibnamefont {Lu}},\ }\bibfield
  {title} {\enquote {\bibinfo {title} {Two-dimensional {Mott} insulators in
  {SrVO}$_3$ ultrathin films},}\ }\href {\doibase 10.1002/admi.201300126}
  {\bibfield  {journal} {\bibinfo  {journal} {Adv. Mater. Interfaces}\ }\textbf
  {\bibinfo {volume} {1}},\ \bibinfo {pages} {1300126} (\bibinfo {year}
  {2014})}\BibitemShut {NoStop}%
\bibitem [{\citenamefont {Bhandary}\ \emph {et~al.}(2016)\citenamefont
  {Bhandary}, \citenamefont {Assmann}, \citenamefont {Aichhorn},\ and\
  \citenamefont {Held}}]{bhandary2016}%
  \BibitemOpen
  \bibfield  {author} {\bibinfo {author} {\bibfnamefont {S.}~\bibnamefont
  {Bhandary}}, \bibinfo {author} {\bibfnamefont {E.}~\bibnamefont {Assmann}},
  \bibinfo {author} {\bibfnamefont {M.}~\bibnamefont {Aichhorn}}, \ and\
  \bibinfo {author} {\bibfnamefont {K.}~\bibnamefont {Held}},\ }\bibfield
  {title} {\enquote {\bibinfo {title} {Charge self-consistency in density
  functional theory combined with dynamical mean field theory: $k$-space
  reoccupation and orbital order},}\ }\href {\doibase
  10.1103/PhysRevB.94.155131} {\bibfield  {journal} {\bibinfo  {journal} {Phys.
  Rev. B}\ }\textbf {\bibinfo {volume} {94}},\ \bibinfo {pages} {155131}
  (\bibinfo {year} {2016})}\BibitemShut {NoStop}%
\bibitem [{\citenamefont {Zhong}\ \emph {et~al.}(2015)\citenamefont {Zhong},
  \citenamefont {Wallerberger}, \citenamefont {Tomczak}, \citenamefont
  {Taranto}, \citenamefont {Parragh}, \citenamefont {Toschi}, \citenamefont
  {Sangiovanni},\ and\ \citenamefont {Held}}]{zhong2015}%
  \BibitemOpen
  \bibfield  {author} {\bibinfo {author} {\bibfnamefont {Z.}~\bibnamefont
  {Zhong}}, \bibinfo {author} {\bibfnamefont {M.}~\bibnamefont {Wallerberger}},
  \bibinfo {author} {\bibfnamefont {J.~M.}\ \bibnamefont {Tomczak}}, \bibinfo
  {author} {\bibfnamefont {C.}~\bibnamefont {Taranto}}, \bibinfo {author}
  {\bibfnamefont {N.}~\bibnamefont {Parragh}}, \bibinfo {author} {\bibfnamefont
  {A.}~\bibnamefont {Toschi}}, \bibinfo {author} {\bibfnamefont
  {G.}~\bibnamefont {Sangiovanni}}, \ and\ \bibinfo {author} {\bibfnamefont
  {K.}~\bibnamefont {Held}},\ }\bibfield  {title} {\enquote {\bibinfo {title}
  {{Electronics with Correlated Oxides: {SrVO$_3$}/{SrTiO$_3$} as a {Mott}
  Transistor}},}\ }\href {\doibase 10.1103/PhysRevLett.114.246401} {\bibfield
  {journal} {\bibinfo  {journal} {Phys. Rev. Lett.}\ }\textbf {\bibinfo
  {volume} {114}},\ \bibinfo {pages} {246401} (\bibinfo {year}
  {2015})}\BibitemShut {NoStop}%
\bibitem [{\citenamefont {Hampel}\ \emph {et~al.}(2020)\citenamefont {Hampel},
  \citenamefont {Beck},\ and\ \citenamefont {Ederer}}]{hampel2020}%
  \BibitemOpen
  \bibfield  {author} {\bibinfo {author} {\bibfnamefont {A.}~\bibnamefont
  {Hampel}}, \bibinfo {author} {\bibfnamefont {S.}~\bibnamefont {Beck}}, \ and\
  \bibinfo {author} {\bibfnamefont {C.}~\bibnamefont {Ederer}},\ }\bibfield
  {title} {\enquote {\bibinfo {title} {Effect of charge self-consistency in
  $\mathrm{DFT}+\mathrm{DMFT}$ calculations for complex transition metal
  oxides},}\ }\href {\doibase 10.1103/PhysRevResearch.2.033088} {\bibfield
  {journal} {\bibinfo  {journal} {Phys. Rev. Research}\ }\textbf {\bibinfo
  {volume} {2}},\ \bibinfo {pages} {033088} (\bibinfo {year}
  {2020})}\BibitemShut {NoStop}%
\bibitem [{\citenamefont {Beck}\ \emph {et~al.}(2018)\citenamefont {Beck},
  \citenamefont {Sclauzero}, \citenamefont {Chopra},\ and\ \citenamefont
  {Ederer}}]{beck2018}%
  \BibitemOpen
  \bibfield  {author} {\bibinfo {author} {\bibfnamefont {S.}~\bibnamefont
  {Beck}}, \bibinfo {author} {\bibfnamefont {G.}~\bibnamefont {Sclauzero}},
  \bibinfo {author} {\bibfnamefont {U.}~\bibnamefont {Chopra}}, \ and\ \bibinfo
  {author} {\bibfnamefont {C.}~\bibnamefont {Ederer}},\ }\bibfield  {title}
  {\enquote {\bibinfo {title} {Metal-insulator transition in {CaVO}$_3$ thin
  films: {I}nterplay between epitaxial strain, dimensional confinement, and
  surface effects},}\ }\href {\doibase 10.1103/PhysRevB.97.075107} {\bibfield
  {journal} {\bibinfo  {journal} {Phys. Rev. B}\ }\textbf {\bibinfo {volume}
  {97}},\ \bibinfo {pages} {075107} (\bibinfo {year} {2018})}\BibitemShut
  {NoStop}%
\bibitem [{\citenamefont {Sclauzero}\ \emph {et~al.}(2016)\citenamefont
  {Sclauzero}, \citenamefont {Dymkowski},\ and\ \citenamefont
  {Ederer}}]{sclauzero2016}%
  \BibitemOpen
  \bibfield  {author} {\bibinfo {author} {\bibfnamefont {G.}~\bibnamefont
  {Sclauzero}}, \bibinfo {author} {\bibfnamefont {K.}~\bibnamefont
  {Dymkowski}}, \ and\ \bibinfo {author} {\bibfnamefont {C.}~\bibnamefont
  {Ederer}},\ }\bibfield  {title} {\enquote {\bibinfo {title} {{Tuning the
  metal-insulator transition in $d^1$ and $d^2$ perovskites by epitaxial
  strain: {A} first-principles-based study}},}\ }\href
  {https://doi.org/10.1103/PhysRevB.94.245109} {\bibfield  {journal} {\bibinfo
  {journal} {Phys. Rev. B}\ }\textbf {\bibinfo {volume} {94}},\ \bibinfo
  {pages} {245109} (\bibinfo {year} {2016})}\BibitemShut {NoStop}%
\bibitem [{\citenamefont {James}\ \emph {et~al.}(2020)\citenamefont {James},
  \citenamefont {Aichhorn},\ and\ \citenamefont {Laverock}}]{james2020quantum}%
  \BibitemOpen
  \bibfield  {author} {\bibinfo {author} {\bibfnamefont {A.~D.~N.}\
  \bibnamefont {James}}, \bibinfo {author} {\bibfnamefont {M.}~\bibnamefont
  {Aichhorn}}, \ and\ \bibinfo {author} {\bibfnamefont {J.}~\bibnamefont
  {Laverock}},\ }\bibfield  {title} {\enquote {\bibinfo {title} {Quantum
  confinement induced metal-insulator transition in strongly correlated quantum
  wells of {SrVO$_3$} superlattices},}\ }\href
  {https://arxiv.org/abs/2005.14329} {\bibfield  {journal} {\bibinfo  {journal}
  {arXiv preprint arXiv:2005.14329}\ } (\bibinfo {year} {2020})}\BibitemShut
  {NoStop}%
\bibitem [{\citenamefont {Blaha}\ \emph {et~al.}(2020)\citenamefont {Blaha},
  \citenamefont {Schwarz}, \citenamefont {Tran}, \citenamefont {Laskowski},
  \citenamefont {Madsen},\ and\ \citenamefont {Marks}}]{blaha2020wien2k}%
  \BibitemOpen
  \bibfield  {author} {\bibinfo {author} {\bibfnamefont {P.}~\bibnamefont
  {Blaha}}, \bibinfo {author} {\bibfnamefont {K.}~\bibnamefont {Schwarz}},
  \bibinfo {author} {\bibfnamefont {F.}~\bibnamefont {Tran}}, \bibinfo {author}
  {\bibfnamefont {R.}~\bibnamefont {Laskowski}}, \bibinfo {author}
  {\bibfnamefont {G.~K.~H.}\ \bibnamefont {Madsen}}, \ and\ \bibinfo {author}
  {\bibfnamefont {L.~D.}\ \bibnamefont {Marks}},\ }\bibfield  {title} {\enquote
  {\bibinfo {title} {{WIEN2k}: An apw+lo program for calculating the properties
  of solids},}\ }\href {\doibase 10.1063/1.5143061} {\bibfield  {journal}
  {\bibinfo  {journal} {The Journal of Chemical Physics}\ }\textbf {\bibinfo
  {volume} {152}},\ \bibinfo {pages} {074101} (\bibinfo {year}
  {2020})}\BibitemShut {NoStop}%
\bibitem [{\citenamefont {Seth}\ \emph {et~al.}(2016)\citenamefont {Seth},
  \citenamefont {Krivenko}, \citenamefont {Ferrero},\ and\ \citenamefont
  {Parcollet}}]{seth2016}%
  \BibitemOpen
  \bibfield  {author} {\bibinfo {author} {\bibfnamefont {P.}~\bibnamefont
  {Seth}}, \bibinfo {author} {\bibfnamefont {I.}~\bibnamefont {Krivenko}},
  \bibinfo {author} {\bibfnamefont {M.}~\bibnamefont {Ferrero}}, \ and\
  \bibinfo {author} {\bibfnamefont {O.}~\bibnamefont {Parcollet}},\ }\bibfield
  {title} {\enquote {\bibinfo {title} {{TRIQS/CTHYB}: A continuous-time quantum
  monte carlo hybridisation expansion solver for quantum impurity problems},}\
  }\href {\doibase 10.1016/j.cpc.2015.10.023} {\bibfield  {journal} {\bibinfo
  {journal} {Comp. Phys. Commun.}\ }\textbf {\bibinfo {volume} {200}},\
  \bibinfo {pages} {274} (\bibinfo {year} {2016})}\BibitemShut {NoStop}%
\bibitem [{\citenamefont {Kraberger}\ and\ \citenamefont {Zingl}()}]{maxent}%
  \BibitemOpen
  \bibfield  {author} {\bibinfo {author} {\bibfnamefont {G.~J.}\ \bibnamefont
  {Kraberger}}\ and\ \bibinfo {author} {\bibfnamefont {M.}~\bibnamefont
  {Zingl}},\ }\href@noop {} {}\bibinfo {howpublished}
  {\url{https://github.com/TRIQS/maxent}}\BibitemShut {NoStop}%
\bibitem [{\citenamefont {Lejaeghere}\ \emph {et~al.}(2016)\citenamefont
  {Lejaeghere}, \citenamefont {Bihlmayer}, \citenamefont {Bj{\"o}rkman},
  \citenamefont {Blaha}, \citenamefont {Bl{\"u}gel}, \citenamefont {Blum},
  \citenamefont {Caliste}, \citenamefont {Castelli}, \citenamefont {Clark},
  \citenamefont {Dal~Corso} \emph {et~al.}}]{Lejaegherea_2016}%
  \BibitemOpen
  \bibfield  {author} {\bibinfo {author} {\bibfnamefont {K.}~\bibnamefont
  {Lejaeghere}}, \bibinfo {author} {\bibfnamefont {G.}~\bibnamefont
  {Bihlmayer}}, \bibinfo {author} {\bibfnamefont {T.}~\bibnamefont
  {Bj{\"o}rkman}}, \bibinfo {author} {\bibfnamefont {P.}~\bibnamefont {Blaha}},
  \bibinfo {author} {\bibfnamefont {S.}~\bibnamefont {Bl{\"u}gel}}, \bibinfo
  {author} {\bibfnamefont {V.}~\bibnamefont {Blum}}, \bibinfo {author}
  {\bibfnamefont {D.}~\bibnamefont {Caliste}}, \bibinfo {author} {\bibfnamefont
  {I.~E.}\ \bibnamefont {Castelli}}, \bibinfo {author} {\bibfnamefont {S.~J.}\
  \bibnamefont {Clark}}, \bibinfo {author} {\bibfnamefont {A.}~\bibnamefont
  {Dal~Corso}},  \emph {et~al.},\ }\bibfield  {title} {\enquote {\bibinfo
  {title} {Reproducibility in density functional theory calculations of
  solids},}\ }\href {\doibase 10.1126/science.aad3000} {\bibfield  {journal}
  {\bibinfo  {journal} {Science}\ }\textbf {\bibinfo {volume} {351}} (\bibinfo
  {year} {2016}),\ 10.1126/science.aad3000}\BibitemShut {NoStop}%
\bibitem [{\citenamefont {Kreyssig}\ \emph {et~al.}(2008)\citenamefont
  {Kreyssig}, \citenamefont {Green}, \citenamefont {Lee}, \citenamefont
  {Samolyuk}, \citenamefont {Zajdel}, \citenamefont {Lynn}, \citenamefont
  {Bud'ko}, \citenamefont {Torikachvili}, \citenamefont {Ni}, \citenamefont
  {Nandi}, \citenamefont {Le\~ao}, \citenamefont {Poulton}, \citenamefont
  {Argyriou}, \citenamefont {Harmon}, \citenamefont {McQueeney}, \citenamefont
  {Canfield},\ and\ \citenamefont {Goldman}}]{Kreyssig2008}%
  \BibitemOpen
  \bibfield  {author} {\bibinfo {author} {\bibfnamefont {A.}~\bibnamefont
  {Kreyssig}}, \bibinfo {author} {\bibfnamefont {M.~A.}\ \bibnamefont {Green}},
  \bibinfo {author} {\bibfnamefont {Y.}~\bibnamefont {Lee}}, \bibinfo {author}
  {\bibfnamefont {G.~D.}\ \bibnamefont {Samolyuk}}, \bibinfo {author}
  {\bibfnamefont {P.}~\bibnamefont {Zajdel}}, \bibinfo {author} {\bibfnamefont
  {J.~W.}\ \bibnamefont {Lynn}}, \bibinfo {author} {\bibfnamefont {S.~L.}\
  \bibnamefont {Bud'ko}}, \bibinfo {author} {\bibfnamefont {M.~S.}\
  \bibnamefont {Torikachvili}}, \bibinfo {author} {\bibfnamefont
  {N.}~\bibnamefont {Ni}}, \bibinfo {author} {\bibfnamefont {S.}~\bibnamefont
  {Nandi}}, \bibinfo {author} {\bibfnamefont {J.~B.}\ \bibnamefont {Le\~ao}},
  \bibinfo {author} {\bibfnamefont {S.~J.}\ \bibnamefont {Poulton}}, \bibinfo
  {author} {\bibfnamefont {D.~N.}\ \bibnamefont {Argyriou}}, \bibinfo {author}
  {\bibfnamefont {B.~N.}\ \bibnamefont {Harmon}}, \bibinfo {author}
  {\bibfnamefont {R.~J.}\ \bibnamefont {McQueeney}}, \bibinfo {author}
  {\bibfnamefont {P.~C.}\ \bibnamefont {Canfield}}, \ and\ \bibinfo {author}
  {\bibfnamefont {A.~I.}\ \bibnamefont {Goldman}},\ }\bibfield  {title}
  {\enquote {\bibinfo {title} {Pressure-induced volume-collapsed tetragonal
  phase of {${\text{CaFe}}_{2}{\text{As}}_{2}$} as seen via neutron
  scattering},}\ }\href {\doibase 10.1103/PhysRevB.78.184517} {\bibfield
  {journal} {\bibinfo  {journal} {Phys. Rev. B}\ }\textbf {\bibinfo {volume}
  {78}},\ \bibinfo {pages} {184517} (\bibinfo {year} {2008})}\BibitemShut
  {NoStop}%
\bibitem [{\citenamefont {{Bud'ko}}\ \emph {et~al.}(2016)\citenamefont
  {{Bud'ko}}, \citenamefont {Ma}, \citenamefont {{Tomi\ifmmode}~{\acute{c}\else
  \'{c}\fi{}}}, \citenamefont {Ran}, \citenamefont {Valent{\'{\i}}},\ and\
  \citenamefont {Canfield}}]{budko2016}%
  \BibitemOpen
  \bibfield  {author} {\bibinfo {author} {\bibfnamefont {S.~L.}\ \bibnamefont
  {{Bud'ko}}}, \bibinfo {author} {\bibfnamefont {X.}~\bibnamefont {Ma}},
  \bibinfo {author} {\bibfnamefont {M.}~\bibnamefont
  {{Tomi\ifmmode}~{\acute{c}\else \'{c}\fi{}}}}, \bibinfo {author}
  {\bibfnamefont {S.}~\bibnamefont {Ran}}, \bibinfo {author} {\bibfnamefont
  {R.}~\bibnamefont {Valent{\'{\i}}}}, \ and\ \bibinfo {author} {\bibfnamefont
  {P.~C.}\ \bibnamefont {Canfield}},\ }\bibfield  {title} {\enquote {\bibinfo
  {title} {Transition to collapsed tetragonal phase in
  {${\mathrm{CaFe}}_{2}{\mathrm{As}}_{2}$} single crystals as seen by
  $^{57}\mathrm{Fe}$ m\"ossbauer spectroscopy},}\ }\href {\doibase
  10.1103/PhysRevB.93.024516} {\bibfield  {journal} {\bibinfo  {journal} {Phys.
  Rev. B}\ }\textbf {\bibinfo {volume} {93}},\ \bibinfo {pages} {024516}
  (\bibinfo {year} {2016})}\BibitemShut {NoStop}%
\bibitem [{\citenamefont {Sypek}\ \emph {et~al.}(2017)\citenamefont {Sypek},
  \citenamefont {Yu}, \citenamefont {Dusoe}, \citenamefont {Drachuck},
  \citenamefont {Patel}, \citenamefont {Giroux}, \citenamefont {Goldman},
  \citenamefont {Kreyssig}, \citenamefont {Canfield}, \citenamefont {Bud’ko}
  \emph {et~al.}}]{sypek2017superelasticity}%
  \BibitemOpen
  \bibfield  {author} {\bibinfo {author} {\bibfnamefont {J.~T.}\ \bibnamefont
  {Sypek}}, \bibinfo {author} {\bibfnamefont {H.}~\bibnamefont {Yu}}, \bibinfo
  {author} {\bibfnamefont {K.~J.}\ \bibnamefont {Dusoe}}, \bibinfo {author}
  {\bibfnamefont {G.}~\bibnamefont {Drachuck}}, \bibinfo {author}
  {\bibfnamefont {H.}~\bibnamefont {Patel}}, \bibinfo {author} {\bibfnamefont
  {A.~M.}\ \bibnamefont {Giroux}}, \bibinfo {author} {\bibfnamefont {A.~I.}\
  \bibnamefont {Goldman}}, \bibinfo {author} {\bibfnamefont {A.}~\bibnamefont
  {Kreyssig}}, \bibinfo {author} {\bibfnamefont {P.~C.}\ \bibnamefont
  {Canfield}}, \bibinfo {author} {\bibfnamefont {S.~L.}\ \bibnamefont
  {Bud’ko}},  \emph {et~al.},\ }\bibfield  {title} {\enquote {\bibinfo
  {title} {Superelasticity and cryogenic linear shape memory effects of
  {CaFe$_2$As$_2$}},}\ }\href {https://doi.org/10.1038/s41467-017-01275-z}
  {\bibfield  {journal} {\bibinfo  {journal} {Nat. Commun.}\ }\textbf {\bibinfo
  {volume} {8}},\ \bibinfo {pages} {1--9} (\bibinfo {year} {2017})}\BibitemShut
  {NoStop}%
\bibitem [{\citenamefont {Saha}\ \emph {et~al.}(2012)\citenamefont {Saha},
  \citenamefont {Butch}, \citenamefont {Drye}, \citenamefont {Magill},
  \citenamefont {Ziemak}, \citenamefont {Kirshenbaum}, \citenamefont {Zavalij},
  \citenamefont {Lynn},\ and\ \citenamefont {Paglione}}]{Saha2012}%
  \BibitemOpen
  \bibfield  {author} {\bibinfo {author} {\bibfnamefont {S.~R.}\ \bibnamefont
  {Saha}}, \bibinfo {author} {\bibfnamefont {N.~P.}\ \bibnamefont {Butch}},
  \bibinfo {author} {\bibfnamefont {T.}~\bibnamefont {Drye}}, \bibinfo {author}
  {\bibfnamefont {J.}~\bibnamefont {Magill}}, \bibinfo {author} {\bibfnamefont
  {S.}~\bibnamefont {Ziemak}}, \bibinfo {author} {\bibfnamefont
  {K.}~\bibnamefont {Kirshenbaum}}, \bibinfo {author} {\bibfnamefont {P.~Y.}\
  \bibnamefont {Zavalij}}, \bibinfo {author} {\bibfnamefont {J.~W.}\
  \bibnamefont {Lynn}}, \ and\ \bibinfo {author} {\bibfnamefont
  {J.}~\bibnamefont {Paglione}},\ }\bibfield  {title} {\enquote {\bibinfo
  {title} {Structural collapse and superconductivity in rare-earth-doped
  {CaFe${}_{2}$As${}_{2}$}},}\ }\href {\doibase 10.1103/PhysRevB.85.024525}
  {\bibfield  {journal} {\bibinfo  {journal} {Phys. Rev. B}\ }\textbf {\bibinfo
  {volume} {85}},\ \bibinfo {pages} {024525} (\bibinfo {year}
  {2012})}\BibitemShut {NoStop}%
\bibitem [{\citenamefont {Kn\"oner}\ \emph {et~al.}(2016)\citenamefont
  {Kn\"oner}, \citenamefont {Gati}, \citenamefont {K\"ohler}, \citenamefont
  {Wolf}, \citenamefont {Tutsch}, \citenamefont {Ran}, \citenamefont
  {Torikachvili}, \citenamefont {Bud'ko}, \citenamefont {Canfield},\ and\
  \citenamefont {Lang}}]{knoner2016}%
  \BibitemOpen
  \bibfield  {author} {\bibinfo {author} {\bibfnamefont {S.}~\bibnamefont
  {Kn\"oner}}, \bibinfo {author} {\bibfnamefont {E.}~\bibnamefont {Gati}},
  \bibinfo {author} {\bibfnamefont {S.}~\bibnamefont {K\"ohler}}, \bibinfo
  {author} {\bibfnamefont {B.}~\bibnamefont {Wolf}}, \bibinfo {author}
  {\bibfnamefont {U.}~\bibnamefont {Tutsch}}, \bibinfo {author} {\bibfnamefont
  {S.}~\bibnamefont {Ran}}, \bibinfo {author} {\bibfnamefont {M.~S.}\
  \bibnamefont {Torikachvili}}, \bibinfo {author} {\bibfnamefont {S.~L.}\
  \bibnamefont {Bud'ko}}, \bibinfo {author} {\bibfnamefont {P.~C.}\
  \bibnamefont {Canfield}}, \ and\ \bibinfo {author} {\bibfnamefont
  {M.}~\bibnamefont {Lang}},\ }\bibfield  {title} {\enquote {\bibinfo {title}
  {Combined effects of {Sr} substitution and pressure on the ground states in
  {${\mathrm{CaFe}}_{2}{\text{As}}_{2}$}},}\ }\href {\doibase
  10.1103/PhysRevB.94.144513} {\bibfield  {journal} {\bibinfo  {journal} {Phys.
  Rev. B}\ }\textbf {\bibinfo {volume} {94}},\ \bibinfo {pages} {144513}
  (\bibinfo {year} {2016})}\BibitemShut {NoStop}%
\bibitem [{\citenamefont {Ran}\ \emph {et~al.}(2014)\citenamefont {Ran},
  \citenamefont {Bud'ko}, \citenamefont {Straszheim},\ and\ \citenamefont
  {Canfield}}]{Ran2014}%
  \BibitemOpen
  \bibfield  {author} {\bibinfo {author} {\bibfnamefont {S.}~\bibnamefont
  {Ran}}, \bibinfo {author} {\bibfnamefont {S.~L.}\ \bibnamefont {Bud'ko}},
  \bibinfo {author} {\bibfnamefont {W.~E.}\ \bibnamefont {Straszheim}}, \ and\
  \bibinfo {author} {\bibfnamefont {P.~C.}\ \bibnamefont {Canfield}},\
  }\bibfield  {title} {\enquote {\bibinfo {title} {Combined effects of
  transition metal ({Ni} and {Rh}) substitution and annealing/quenching on the
  physical properties of {${\mathrm{CaFe}}_{2}{\mathrm{As}}_{2}$}},}\ }\href
  {\doibase 10.1103/PhysRevB.90.054501} {\bibfield  {journal} {\bibinfo
  {journal} {Phys. Rev. B}\ }\textbf {\bibinfo {volume} {90}},\ \bibinfo
  {pages} {054501} (\bibinfo {year} {2014})}\BibitemShut {NoStop}%
\bibitem [{\citenamefont {Nohara}\ and\ \citenamefont
  {Kudo}(2017)}]{Nohara2017}%
  \BibitemOpen
  \bibfield  {author} {\bibinfo {author} {\bibfnamefont {M.}~\bibnamefont
  {Nohara}}\ and\ \bibinfo {author} {\bibfnamefont {K.}~\bibnamefont {Kudo}},\
  }\bibfield  {title} {\enquote {\bibinfo {title} {Arsenic chemistry of
  iron-based superconductors and strategy for novel superconducting
  materials},}\ }\href {\doibase 10.1080/23746149.2017.1317024} {\bibfield
  {journal} {\bibinfo  {journal} {Advances in Physics: X}\ }\textbf {\bibinfo
  {volume} {2}},\ \bibinfo {pages} {450--461} (\bibinfo {year}
  {2017})}\BibitemShut {NoStop}%
\bibitem [{\citenamefont {Pobel}\ \emph {et~al.}(2013)\citenamefont {Pobel},
  \citenamefont {Frankovsky},\ and\ \citenamefont {Johrendt}}]{Pobel2013}%
  \BibitemOpen
  \bibfield  {author} {\bibinfo {author} {\bibfnamefont {R.}~\bibnamefont
  {Pobel}}, \bibinfo {author} {\bibfnamefont {R.}~\bibnamefont {Frankovsky}}, \
  and\ \bibinfo {author} {\bibfnamefont {D.}~\bibnamefont {Johrendt}},\
  }\bibfield  {title} {\enquote {\bibinfo {title} {Ferromagnetism and the
  formation of interlayer {As2} dimers in {Ca(Fe$_{\rm 1-x}$Ni$_{\rm
  x}$)2As$_2$}},}\ }\href {\doibase https://doi.org/10.5560/znb.2013-3045}
  {\bibfield  {journal} {\bibinfo  {journal} {Zeitschrift für Naturforschung
  B}\ }\textbf {\bibinfo {volume} {68}},\ \bibinfo {pages} {581 -- 586}
  (\bibinfo {year} {2013})}\BibitemShut {NoStop}%
\bibitem [{\citenamefont {Ortenzi}\ \emph {et~al.}(2015)\citenamefont
  {Ortenzi}, \citenamefont {Gretarsson}, \citenamefont {Kasahara},
  \citenamefont {Matsuda}, \citenamefont {Shibauchi}, \citenamefont
  {Finkelstein}, \citenamefont {Wu}, \citenamefont {Julian}, \citenamefont
  {Kim}, \citenamefont {Mazin},\ and\ \citenamefont {Boeri}}]{Ortenzi2015}%
  \BibitemOpen
  \bibfield  {author} {\bibinfo {author} {\bibfnamefont {L.}~\bibnamefont
  {Ortenzi}}, \bibinfo {author} {\bibfnamefont {H.}~\bibnamefont {Gretarsson}},
  \bibinfo {author} {\bibfnamefont {S.}~\bibnamefont {Kasahara}}, \bibinfo
  {author} {\bibfnamefont {Y.}~\bibnamefont {Matsuda}}, \bibinfo {author}
  {\bibfnamefont {T.}~\bibnamefont {Shibauchi}}, \bibinfo {author}
  {\bibfnamefont {K.~D.}\ \bibnamefont {Finkelstein}}, \bibinfo {author}
  {\bibfnamefont {W.}~\bibnamefont {Wu}}, \bibinfo {author} {\bibfnamefont
  {S.~R.}\ \bibnamefont {Julian}}, \bibinfo {author} {\bibfnamefont
  {Young-June}\ \bibnamefont {Kim}}, \bibinfo {author} {\bibfnamefont {I.~I.}\
  \bibnamefont {Mazin}}, \ and\ \bibinfo {author} {\bibfnamefont
  {L.}~\bibnamefont {Boeri}},\ }\bibfield  {title} {\enquote {\bibinfo {title}
  {Structural origin of the anomalous temperature dependence of the local
  magnetic moments in the {${\mathrm{CaFe}}_{2}{\mathrm{As}}_{2}$} family of
  materials},}\ }\href {\doibase 10.1103/PhysRevLett.114.047001} {\bibfield
  {journal} {\bibinfo  {journal} {Phys. Rev. Lett.}\ }\textbf {\bibinfo
  {volume} {114}},\ \bibinfo {pages} {047001} (\bibinfo {year}
  {2015})}\BibitemShut {NoStop}%
\bibitem [{\citenamefont {Yang}\ \emph {et~al.}(2015)\citenamefont {Yang},
  \citenamefont {Le}, \citenamefont {Zhang}, \citenamefont {Xu}, \citenamefont
  {Zhang}, \citenamefont {Nadeem}, \citenamefont {Xiao}, \citenamefont {Hu},\
  and\ \citenamefont {Qiu}}]{Yang2015}%
  \BibitemOpen
  \bibfield  {author} {\bibinfo {author} {\bibfnamefont {R.}~\bibnamefont
  {Yang}}, \bibinfo {author} {\bibfnamefont {C.}~\bibnamefont {Le}}, \bibinfo
  {author} {\bibfnamefont {L.}~\bibnamefont {Zhang}}, \bibinfo {author}
  {\bibfnamefont {B.}~\bibnamefont {Xu}}, \bibinfo {author} {\bibfnamefont
  {W.}~\bibnamefont {Zhang}}, \bibinfo {author} {\bibfnamefont
  {K.}~\bibnamefont {Nadeem}}, \bibinfo {author} {\bibfnamefont
  {H.}~\bibnamefont {Xiao}}, \bibinfo {author} {\bibfnamefont {J.}~\bibnamefont
  {Hu}}, \ and\ \bibinfo {author} {\bibfnamefont {X.}~\bibnamefont {Qiu}},\
  }\bibfield  {title} {\enquote {\bibinfo {title} {Formation of {As-As} bond
  and its effect on absence of superconductivity in the collapsed tetragonal
  phase of
  {${\mathrm{Ca}}_{0.86}{\mathrm{Pr}}_{0.14}{\mathrm{Fe}}_{2}{\mathrm{As}}_{2}$}:
  An optical spectroscopy study},}\ }\href {\doibase
  10.1103/PhysRevB.91.224507} {\bibfield  {journal} {\bibinfo  {journal} {Phys.
  Rev. B}\ }\textbf {\bibinfo {volume} {91}},\ \bibinfo {pages} {224507}
  (\bibinfo {year} {2015})}\BibitemShut {NoStop}%
\bibitem [{\citenamefont {van Roekeghem}\ \emph {et~al.}(2016)\citenamefont
  {van Roekeghem}, \citenamefont {Richard}, \citenamefont {Shi}, \citenamefont
  {Wu}, \citenamefont {Zeng}, \citenamefont {Saparov}, \citenamefont {Ohtsubo},
  \citenamefont {Qian}, \citenamefont {Sefat}, \citenamefont {Biermann},\ and\
  \citenamefont {Ding}}]{vanroekeghem2016}%
  \BibitemOpen
  \bibfield  {author} {\bibinfo {author} {\bibfnamefont {A.}~\bibnamefont {van
  Roekeghem}}, \bibinfo {author} {\bibfnamefont {P.}~\bibnamefont {Richard}},
  \bibinfo {author} {\bibfnamefont {X.}~\bibnamefont {Shi}}, \bibinfo {author}
  {\bibfnamefont {S.}~\bibnamefont {Wu}}, \bibinfo {author} {\bibfnamefont
  {L.}~\bibnamefont {Zeng}}, \bibinfo {author} {\bibfnamefont {B.}~\bibnamefont
  {Saparov}}, \bibinfo {author} {\bibfnamefont {Y.}~\bibnamefont {Ohtsubo}},
  \bibinfo {author} {\bibfnamefont {T.}~\bibnamefont {Qian}}, \bibinfo {author}
  {\bibfnamefont {A.~S.}\ \bibnamefont {Sefat}}, \bibinfo {author}
  {\bibfnamefont {S.}~\bibnamefont {Biermann}}, \ and\ \bibinfo {author}
  {\bibfnamefont {H.}~\bibnamefont {Ding}},\ }\bibfield  {title} {\enquote
  {\bibinfo {title} {Tetragonal and collapsed-tetragonal phases of
  {${\mathrm{CaFe}}_{2}{\mathrm{As}}_{2}$}: A view from angle-resolved
  photoemission and dynamical mean-field theory},}\ }\href {\doibase
  10.1103/PhysRevB.93.245139} {\bibfield  {journal} {\bibinfo  {journal} {Phys.
  Rev. B}\ }\textbf {\bibinfo {volume} {93}},\ \bibinfo {pages} {245139}
  (\bibinfo {year} {2016})}\BibitemShut {NoStop}%
\bibitem [{\citenamefont {Dhaka}\ \emph {et~al.}(2014)\citenamefont {Dhaka},
  \citenamefont {Jiang}, \citenamefont {Ran}, \citenamefont {{Bud'ko}},
  \citenamefont {Canfield}, \citenamefont {Harmon}, \citenamefont {Kaminski},
  \citenamefont {{Tomi\ifmmode \acute{c}\else \'{c}\fi{}}}, \citenamefont
  {{Valent\'{\i}}},\ and\ \citenamefont {Lee}}]{dhaka2014}%
  \BibitemOpen
  \bibfield  {author} {\bibinfo {author} {\bibfnamefont {R.~S.}\ \bibnamefont
  {Dhaka}}, \bibinfo {author} {\bibfnamefont {R.}~\bibnamefont {Jiang}},
  \bibinfo {author} {\bibfnamefont {S.}~\bibnamefont {Ran}}, \bibinfo {author}
  {\bibfnamefont {S.~L.}\ \bibnamefont {{Bud'ko}}}, \bibinfo {author}
  {\bibfnamefont {P.~C.}\ \bibnamefont {Canfield}}, \bibinfo {author}
  {\bibfnamefont {B.~N.}\ \bibnamefont {Harmon}}, \bibinfo {author}
  {\bibfnamefont {A.}~\bibnamefont {Kaminski}}, \bibinfo {author}
  {\bibfnamefont {M.}~\bibnamefont {{Tomi\ifmmode \acute{c}\else \'{c}\fi{}}}},
  \bibinfo {author} {\bibfnamefont {R.}~\bibnamefont {{Valent\'{\i}}}}, \ and\
  \bibinfo {author} {\bibfnamefont {Y.}~\bibnamefont {Lee}},\ }\bibfield
  {title} {\enquote {\bibinfo {title} {Dramatic changes in the electronic
  structure upon transition to the collapsed tetragonal phase in
  {CaFe${}_{2}$As${}_{2}$}},}\ }\href {\doibase 10.1103/PhysRevB.89.020511}
  {\bibfield  {journal} {\bibinfo  {journal} {Phys. Rev. B}\ }\textbf {\bibinfo
  {volume} {89}},\ \bibinfo {pages} {020511(R)} (\bibinfo {year}
  {2014})}\BibitemShut {NoStop}%
\bibitem [{\citenamefont {Diehl}\ \emph {et~al.}(2014)\citenamefont {Diehl},
  \citenamefont {Backes}, \citenamefont {Guterding}, \citenamefont {Jeschke},\
  and\ \citenamefont {Valent{\'\i}}}]{diehl2014correlation}%
  \BibitemOpen
  \bibfield  {author} {\bibinfo {author} {\bibfnamefont {J.}~\bibnamefont
  {Diehl}}, \bibinfo {author} {\bibfnamefont {S.}~\bibnamefont {Backes}},
  \bibinfo {author} {\bibfnamefont {D.}~\bibnamefont {Guterding}}, \bibinfo
  {author} {\bibfnamefont {H.~O.}\ \bibnamefont {Jeschke}}, \ and\ \bibinfo
  {author} {\bibfnamefont {R.}~\bibnamefont {Valent{\'\i}}},\ }\bibfield
  {title} {\enquote {\bibinfo {title} {Correlation effects in the tetragonal
  and collapsed-tetragonal phase of {CaFe$_2$As$_2$}},}\ }\href
  {https://journals.aps.org/prb/pdf/10.1103/PhysRevB.90.085110} {\bibfield
  {journal} {\bibinfo  {journal} {Phys. Rev. B}\ }\textbf {\bibinfo {volume}
  {90}},\ \bibinfo {pages} {085110} (\bibinfo {year} {2014})}\BibitemShut
  {NoStop}%
\bibitem [{\citenamefont {Mandal}\ \emph {et~al.}(2014)\citenamefont {Mandal},
  \citenamefont {Cohen},\ and\ \citenamefont {Haule}}]{Mandal2014}%
  \BibitemOpen
  \bibfield  {author} {\bibinfo {author} {\bibfnamefont {S.}~\bibnamefont
  {Mandal}}, \bibinfo {author} {\bibfnamefont {R.~E.}\ \bibnamefont {Cohen}}, \
  and\ \bibinfo {author} {\bibfnamefont {K.}~\bibnamefont {Haule}},\ }\bibfield
   {title} {\enquote {\bibinfo {title} {Pressure suppression of electron
  correlation in the collapsed tetragonal phase of
  {${\mathrm{CaFe}}_{2}{\mathrm{As}}_{2}$}: A {DFT-DMFT} investigation},}\
  }\href {\doibase 10.1103/PhysRevB.90.060501} {\bibfield  {journal} {\bibinfo
  {journal} {Phys. Rev. B}\ }\textbf {\bibinfo {volume} {90}},\ \bibinfo
  {pages} {060501(R)} (\bibinfo {year} {2014})}\BibitemShut {NoStop}%
\bibitem [{\citenamefont {Hoffmann}\ and\ \citenamefont
  {Zheng}(1985)}]{hoffmann1985}%
  \BibitemOpen
  \bibfield  {author} {\bibinfo {author} {\bibfnamefont {R.}~\bibnamefont
  {Hoffmann}}\ and\ \bibinfo {author} {\bibfnamefont {C.}~\bibnamefont
  {Zheng}},\ }\bibfield  {title} {\enquote {\bibinfo {title} {Making and
  breaking bonds in the solid state: the {ThCr$_2$Si$_2$} structure},}\ }\href
  {https://pubs.acs.org/doi/abs/10.1021/j100266a007} {\bibfield  {journal}
  {\bibinfo  {journal} {The Journal of Physical Chemistry}\ }\textbf {\bibinfo
  {volume} {89}},\ \bibinfo {pages} {4175--4181} (\bibinfo {year}
  {1985})}\BibitemShut {NoStop}%
\bibitem [{\citenamefont {Zheng}\ and\ \citenamefont
  {Hoffmann}(1988)}]{zheng1988}%
  \BibitemOpen
  \bibfield  {author} {\bibinfo {author} {\bibfnamefont {C.}~\bibnamefont
  {Zheng}}\ and\ \bibinfo {author} {\bibfnamefont {R.}~\bibnamefont
  {Hoffmann}},\ }\bibfield  {title} {\enquote {\bibinfo {title} {Complementary
  local and extended views of bonding in the {ThCr$_2$Si$_2$} and
  {CaAl$_2$Si$_2$} structures},}\ }\href
  {https://doi.org/10.1016/0022-4596(88)90009-6} {\bibfield  {journal}
  {\bibinfo  {journal} {Journal of Solid State Chemistry}\ }\textbf {\bibinfo
  {volume} {72}},\ \bibinfo {pages} {58--71} (\bibinfo {year}
  {1988})}\BibitemShut {NoStop}%
\bibitem [{\citenamefont {Stavrou}\ \emph {et~al.}(2015)\citenamefont
  {Stavrou}, \citenamefont {Chen}, \citenamefont {Oganov}, \citenamefont
  {Wang}, \citenamefont {Yan}, \citenamefont {Luo}, \citenamefont {Chen},\ and\
  \citenamefont {Goncharov}}]{stavrou2015}%
  \BibitemOpen
  \bibfield  {author} {\bibinfo {author} {\bibfnamefont {E.}~\bibnamefont
  {Stavrou}}, \bibinfo {author} {\bibfnamefont {X.}~\bibnamefont {Chen}},
  \bibinfo {author} {\bibfnamefont {A.~R.}\ \bibnamefont {Oganov}}, \bibinfo
  {author} {\bibfnamefont {A.~F.}\ \bibnamefont {Wang}}, \bibinfo {author}
  {\bibfnamefont {Y.~J.}\ \bibnamefont {Yan}}, \bibinfo {author} {\bibfnamefont
  {X.~G.}\ \bibnamefont {Luo}}, \bibinfo {author} {\bibfnamefont {X.~H.}\
  \bibnamefont {Chen}}, \ and\ \bibinfo {author} {\bibfnamefont {A.~F.}\
  \bibnamefont {Goncharov}},\ }\bibfield  {title} {\enquote {\bibinfo {title}
  {Formation of {As-As} interlayer bonding in the collapsed tetragonal phase of
  {NaFe$_2$As$_2$} under pressure},}\ }\href
  {https://doi.org/10.1038/srep09868} {\bibfield  {journal} {\bibinfo
  {journal} {Scientific Reports}\ }\textbf {\bibinfo {volume} {5}},\ \bibinfo
  {pages} {9868} (\bibinfo {year} {2015})}\BibitemShut {NoStop}%
\bibitem [{\citenamefont {Yildirim}(2009)}]{Yildirim2009}%
  \BibitemOpen
  \bibfield  {author} {\bibinfo {author} {\bibfnamefont {T.}~\bibnamefont
  {Yildirim}},\ }\bibfield  {title} {\enquote {\bibinfo {title} {Strong
  coupling of the {Fe}-spin state and the {As-As} hybridization in
  iron-pnictide superconductors from first-principle calculations},}\ }\href
  {\doibase 10.1103/PhysRevLett.102.037003} {\bibfield  {journal} {\bibinfo
  {journal} {Phys. Rev. Lett.}\ }\textbf {\bibinfo {volume} {102}},\ \bibinfo
  {pages} {037003} (\bibinfo {year} {2009})}\BibitemShut {NoStop}%
\bibitem [{\citenamefont {Peil}\ \emph {et~al.}(2019)\citenamefont {Peil},
  \citenamefont {Hampel}, \citenamefont {Ederer},\ and\ \citenamefont
  {Georges}}]{Peil_2019}%
  \BibitemOpen
  \bibfield  {author} {\bibinfo {author} {\bibfnamefont {O.~E.}\ \bibnamefont
  {Peil}}, \bibinfo {author} {\bibfnamefont {A.}~\bibnamefont {Hampel}},
  \bibinfo {author} {\bibfnamefont {C.}~\bibnamefont {Ederer}}, \ and\ \bibinfo
  {author} {\bibfnamefont {A.}~\bibnamefont {Georges}},\ }\bibfield  {title}
  {\enquote {\bibinfo {title} {Mechanism and control parameters of the coupled
  structural and metal-insulator transition in nickelates},}\ }\href {\doibase
  10.1103/PhysRevB.99.245127} {\bibfield  {journal} {\bibinfo  {journal} {Phys.
  Rev. B}\ }\textbf {\bibinfo {volume} {99}},\ \bibinfo {pages} {245127}
  (\bibinfo {year} {2019})}\BibitemShut {NoStop}%
\bibitem [{\citenamefont {Cooper}\ \emph {et~al.}(2004)\citenamefont {Cooper},
  \citenamefont {Mijnarends}, \citenamefont {Shiotani}, \citenamefont {Sakai},\
  and\ \citenamefont {Bansil}}]{cooper2004x}%
  \BibitemOpen
  \bibfield  {author} {\bibinfo {author} {\bibfnamefont {M.~J.}\ \bibnamefont
  {Cooper}}, \bibinfo {author} {\bibfnamefont {P.~E.}\ \bibnamefont
  {Mijnarends}}, \bibinfo {author} {\bibfnamefont {N.}~\bibnamefont
  {Shiotani}}, \bibinfo {author} {\bibfnamefont {N.}~\bibnamefont {Sakai}}, \
  and\ \bibinfo {author} {\bibfnamefont {A.}~\bibnamefont {Bansil}},\
  }\href@noop {} {\emph {\bibinfo {title} {X-Ray Compton Scattering}}},\ Oxford
  Series on Synchrotron Radiation\ (\bibinfo  {publisher} {OUP Oxford},\
  \bibinfo {year} {2004})\BibitemShut {NoStop}%
\bibitem [{\citenamefont {Sharma}\ \emph {et~al.}(2014)\citenamefont {Sharma},
  \citenamefont {Dewhurst},\ and\ \citenamefont {Gross}}]{Sharma_2014}%
  \BibitemOpen
  \bibfield  {author} {\bibinfo {author} {\bibfnamefont {S.}~\bibnamefont
  {Sharma}}, \bibinfo {author} {\bibfnamefont {J.~K.}\ \bibnamefont
  {Dewhurst}}, \ and\ \bibinfo {author} {\bibfnamefont {E.~K.~U.}\ \bibnamefont
  {Gross}},\ }\enquote {\bibinfo {title} {Optical response of extended systems
  using time-dependent density functional theory},}\ \ (\bibinfo  {publisher}
  {Springer},\ \bibinfo {year} {2014})\BibitemShut {NoStop}%
\bibitem [{\citenamefont {Galicia-Hernandez}\ \emph {et~al.}(2020)\citenamefont
  {Galicia-Hernandez}, \citenamefont {Turkowski}, \citenamefont
  {Hernandez-Cocoletzi},\ and\ \citenamefont
  {Rahman}}]{Galicia_Hernandez_2020}%
  \BibitemOpen
  \bibfield  {author} {\bibinfo {author} {\bibfnamefont {J.~M.}\ \bibnamefont
  {Galicia-Hernandez}}, \bibinfo {author} {\bibfnamefont {V.}~\bibnamefont
  {Turkowski}}, \bibinfo {author} {\bibfnamefont {G.}~\bibnamefont
  {Hernandez-Cocoletzi}}, \ and\ \bibinfo {author} {\bibfnamefont {T.~S.}\
  \bibnamefont {Rahman}},\ }\bibfield  {title} {\enquote {\bibinfo {title}
  {Electron correlations and memory effects in ultrafast electron and hole
  dynamics in {VO}$_2$},}\ }\href {\doibase 10.1088/1361-648x/ab6f85}
  {\bibfield  {journal} {\bibinfo  {journal} {Journal of Physics: Condensed
  Matter}\ }\textbf {\bibinfo {volume} {32}},\ \bibinfo {pages} {20LT01}
  (\bibinfo {year} {2020})}\BibitemShut {NoStop}%
\end{thebibliography}
\end{document}